\RequirePackage{fix-cm}
\documentclass{svjour3}                     
\smartqed  
\usepackage{graphicx}
\usepackage{txfonts}
\usepackage{siunitx}
\usepackage{enumerate}
\usepackage{hyperref}

 \newcommand{\change}[1]{{\rm #1}}
\journalname{Experimental Astronomy}
\begin{document}

\title{The CHEOPS mission}



\author{
W. Benz$^{1 \& 2}$,
C. Broeg$^2$,
A. Fortier$^2$, 
N. Rando$^3$, 
T. Beck$^2$,  
M. Beck$^4$, 
D. Queloz$^{4 \& 5}$, 
D. Ehrenreich$^4$, 
P.F.L. Maxted$^6$, 
K.G. Isaak$^3$
% ...
}

\authorrunning{W. Benz et al.} 

\institute{
$^1$ Physikalisches Institut, University of Bern, Switzerland
\\
$^2$ Center for Space and Habitability, University of Bern, Switzerland
\\
$^3$ ESTEC, European Space Agency, Noordwijk, NL 
\\
$^4$ Observatoire de Gen\`eve, University of Geneva, Switzerland  
\\
$^{5}$ Institute of Astronomy, University of Cambridge, UK 
\\
$^6$ Astrophysics Group, Keele University, Keele University, Staffordshire, UK
\\
$^7$ Instituto de Astrof\'\i sica de Canarias, La Laguna, Tenerife, Spain
\\
$^8$ Dpto. de Astrof\'isica, Universidad de La Laguna, Tenerife, Spain
\\
$^{9}$ DEIMOS Engenharia, Lisboa, Portugal
\\
$^{10}$ Admatis, Miskolc, Hungary 
\\
$^{11}$ Unidad de Excelencia Mar\'{i}a de Maeztu- Centro de Astrobiolog\'{i}a (INTA-CSIC) 
\\
$^{12}$ Instituto de Astrof\'isica e Ci\^encias do Espa\c{c}o, Universidade do Porto, CAUP, Portugal
\\
$^{13}$ Departamento de F\'isica e Astronomia, Faculdade de Ci\^encias, Universidade do Porto, Portugal
\\
$^{14}$ Space Research Institute, Austrian Academy of Sciences, Graz, Austria 
\\
$^{15}$ INAF Astronomical Observatory of Padova, Italy 
\\
$^{16}$ Institut de Plan\'etologie et d'Astrophysique, University of Grenoble, France 
\\
$^{17}$ Stockholm University, Sweden 
\\
$^{18}$ Institute of Planetary Research, German Aerospace Center (DLR), Berlin, Germany 
\\
$^{19}$ Institut de Physique du Globe, Paris, France 
\\
$^{20}$ University of St. Andrews, UK 
\\
$^{21}$ AIRBUS Defence \& Space Earth Observation, Navigation \& Science, Madrid, Spain 
\\
$^{22}$ Lund Observatory, Lund University, Sweden
\\
$^{23}$ Aix Marseille Univ, CNRS, CNES, LAM, Marseille, France 
\\
$^{24}$ Space Sciences, Technologies and Astrophysics Research Institute, Universit\'{e} de Li\`{e}ge, Belgium
\\
$^{25}$ Astrobiology Research Unit, Universit\'{e} de Li\`{e}ge, Belgium
\\
$^{26}$ University of Leiden, The Netherlands
\\
$^{27}$ Chalmers University, G\"{o}teborg, Sweden
\\
$^{28}$ Universita degli Studi di Torino, University of Torino, Italy
\\
$^{29}$ Zentrum f\"ur Astronomie und Astrophysik, Technische Universit\"at Berlin, Germany
\\
$^{30}$ Division Technique INSU, La Seyne sur Mer, France
\\
$^{31}$ Konkoly Observatory, Research Centre for Astronomy and Earth Sciences, Budapest, Hungary
\\
$^{32}$ ELTE E\"otv\"os Lor\'and University, Institute of Physics, P\'azm\'any P\'eter s\'et\'any 1/A, Budapest, Hungary
\\
$^{33}$ IMCCE, CNRS, Observatoire de Paris, PSL University, Sorbonne Université,  France 
\\
$^{34}$ Institut d'Astrophysique de Paris, France 
\\
$^{35}$ INTA, Instituto Nacional de T\'{e}cnica Aeroespacial, Torrej\`{o}n de Ardoz, Madrid, Spain
\\
$^{36}$ Department of Astrophysics, University of Vienna, Austria 
\\
$^{37}$ INAF Astrophysical Observatory of Catania, Italy 
\\
$^{38}$ Department of Physics and Astronomy, University of Padova, Italy
\\
$^{39}$ Institute of Optical Sensor Systems, German Aerospace Center (DLR), Berlin, Germany
\\
$^{40}$ Department of Physics, University of Warwick, Coventry, UK
\\
$^{41}$ Institute of Geological Sciences, FU Berlin, Germany
\\
$^{42}$ Institut de Ci\`{e}nces de l'Espai (ICE, CSIC), Campus UAB, Bellaterra, Spain
\\
$^{43}$ Institut d'Estudis Espacials de Catalunya (IEEC), Barcelona, Spain
\\
$^{44}$ Italian Space Agency, Rome, Italy
\\
$^{45}$ ELTE E\"otv\"os Lor\'and University, Gothard Astrophysical Observatory, Szombathely, Hungary
\\
$^{46}$ MTA-ELTE Exoplanet Research Group, Szombathely, Szent Imre h. u. 112, Hungary
}

\date{Received: date / Accepted: date}

\maketitle

\begin{abstract}

The CHaracterising ExOPlanet Satellite (CHEOPS) was selected on October 19, 2012, as the first small mission (S-mission) in the ESA Science Programme and successfully launched on December 18, 2019, as a secondary passenger on a Soyuz-Fregat rocket from Kourou, French Guiana. CHEOPS is a partnership between ESA and Switzerland with important contributions by ten additional ESA Member States. CHEOPS is the first mission dedicated to search for transits of exoplanets using ultrahigh precision photometry on bright stars already known to host planets. As a follow-up mission, CHEOPS is mainly dedicated to improving, whenever possible, existing radii measurements or provide first accurate measurements for a subset of those planets for which the mass has already been estimated from ground-based spectroscopic surveys. The expected photometric precision will also allow CHEOPS to go beyond measuring only transits and to follow phase curves or to search for exo-moons, for example. Finally, by unveiling transiting exoplanets with high potential for in-depth characterisation, CHEOPS will also provide prime targets for future instruments suited to the spectroscopic characterisation of exoplanetary atmospheres.

To reach its science objectives, requirements on the photometric precision and stability have been derived for stars with magnitudes ranging from 6 to 12 in the V band. In particular, CHEOPS shall be able to detect Earth-size planets transiting G5 dwarf stars (stellar radius of \(0.9\,R_{\odot}\)) in the magnitude range $6 \le V \le 9$ by achieving a photometric precision of \SI{20} {ppm} in 6 hours of integration time. In the case of K-type stars (stellar radius of \(0.7\,R_{\odot}\)) of magnitude in the range $9 \le V \le 12$, CHEOPS shall be able to detect transiting Neptune-size planets achieving a photometric precision of \SI{85}{ppm} in 3 hours of integration time. This precision has to be maintained over continuous periods of observation for up to 48 hours. This precision and stability will be achieved by using a single, frame-transfer, back-illuminated CCD detector at the focal plane assembly of a \SI{33.5}{cm} diameter, on-axis Ritchey-Chr\'{e}tien telescope. The nearly \SI{280}{kg} spacecraft is nadir-locked, with a pointing accuracy of about 1 arcsec rms, and will allow for at least 1 Gbit/day downlink. The sun-synchronous dusk-dawn orbit at 700 km altitude enables having the Sun permanently on the backside of the spacecraft thus minimising Earth stray light. 

A mission duration of \SI{3.5}{years} in orbit is foreseen to enable the execution of the science programme. During this period, 20\% of the observing time is available to the wider community through yearly ESA call for proposals, as well as through discretionary time approved by ESA's Director of Science. At the time of this writing, CHEOPS commissioning has been completed and CHEOPS has been shown to fulfill all its requirements. The mission has now started the execution of its science programme.  
\end{abstract}

\section{Introduction}
\label{intro}
In March 2012, the European Space Agency (ESA) issued a call for a small mission opportunity. This new class of mission (S-class) in the portfolio of the science programme of the Agency was introduced to provide the community with additional launch opportunities without interfering with the existing programme based on M- and L-class missions. This translated into strict boundary conditions on the potential mission(s). The selection would be based on scientific excellence, the possibility for fast development, and a total cost to the Agency, including launch, not exceeding 50 MEuros. 

The CHEOPS (CHaracterising ExOPlanet Satellite)\footnote{\url{https://https://cheops.unibe.ch/}} proposal was submitted in response to the call by a Consortium of research institutes located in ESA member states. It was subsequently selected by ESA’s Space Programme Committee in November 2012 out of a total of 26 competing submissions. 

Following the discovery in 1995 of the first exoplanet \cite{mayorqueloz:1995} orbiting a solar-like star and the subsequent detection of over 4000 additional planets, the need for physical and chemical characterisation of these objects is growing. For this, a sample as large as possible of planets orbiting bright stars and for which mass \textit{and} radius are precisely measured is needed. This need triggered in 2008 the idea of developing a space mission dedicated to searching for exoplanetary transits by performing ultra-high precision photometry on bright stars already known to host planets. Subsequently, a study was conducted in Switzerland in 2010-2011 which concluded that such a concept was indeed feasible but that it would exceed the financial capabilities of Switzerland. A Consortium was built and a proposal submitted in response to ESA's call from small missions. 

 The follow-up nature of CHEOPS, with a single star being targeted at a time, makes this transit mission unique compared to its successful precursors COROT \cite{Baglin:2006}, Kepler \cite{Koch:2010}, and TESS \cite{Ricker:2014}, or its successor PLATO \cite{Rauer:2014}. This difference is the basis for an original science programme (see Sect.~\ref{sec:science}) in which the focus is not on the discovery of additional exoplanets, but rather on the characterisation of a set of most promising objects for constraining planet formation and evolution theories and for further studies by future large infrastructures (e.g. JWST, Ariel, ELTs). This resulted in several challenging requirements on photometric precision and sky visibility (see Sect.~\ref{sec:requirements}) which drove the design of the mission. With its unique characteristics, CHEOPS is complementary to all other transit missions as it provides the agility and the photometric precision necessary to re-visit sufficiently interesting targets for which further measurements are deemed essential \cite{Benz:2018}. 

The boundary conditions imposed by ESA on the S-missions briefly mentioned above (see Sect.~\ref{sec:implementation}) translated into several trade-off decisions being taken while selecting and adapting the platform (see Sect.~\ref{sec:mission_SC_design}), designing the payload (see Sect.~\ref{sec:CHEOPS_PAYLOAD}), and developing the ground segment (see Sect.~\ref{sec:Ground_segment}). In this sense, CHEOPS is possibly not the ultimate follow-up mission that could have been flown (if such a thing exists), but it is arguably the best that could be built within the ESA framework under the conditions given. In the laboratory, CHEOPS has met or exceeded all the specifications that could be measured. It was then successfully launched on 18 December 2019 from Kourou, French Guyana, as a secondary passenger on a Soyuz-Fregat rocket. Commissioning ended in March 2020 demonstrating that CHEOPS is meeting all the requirements providing the green light for the science phase. With the successful completion of commissioning, CHEOPS has not only met all the technical requirements but also the programmatic ones namely the overall cost and schedule. A remarkable achievement. 

\section{The CHEOPS Science Programme}
\label{sec:science}
The science observing time on CHEOPS is foreseen to fill a minimum of 90\% of the nominal lifetime of the mission following successful commissioning of the satellite, with the remaining 10\% split between activities related to spacecraft operations (eg. safe mode and recovery, anomaly investigation, instrument software update, etc.), and a monitoring and characterisation campaign designed to monitor the performances of the CHEOPS instrument throughout the mission. The underlying rules on how the science time is to be used are defined in the CHEOPS Science Management Plan, which foresees an 80\%:20\%  split between the Core or Guaranteed Time Observing (GTO) Programme that is under the responsibility of the CHEOPS Science Team (see \ref{subsec:science_GTO}), and the Guest Observers (GO) Programme (see \ref{subsec:science_GO}), under the responsibility of ESA and through which the community can conduct investigations of their choice. 

The CHEOPS Science Team defines the first, prioritised target list covering observations for the GTO programme for up to the duration of the nominal mission in advance of the definition of the GO programme. These targets are placed on a reserved list that cannot be observed as part of the GO Programme. An update of the reserved target list is allowed throughout the mission, with some restrictions. The GO programme is then built out of annual calls issued by ESA. The targets proposed and approved by an ESA appointed Time Allocating Committee are added to the reserved target list and can only be observed by the proposing team. 

\subsection{The Guaranteed Time Observing  Programme Science}
\label{subsec:science_GTO}
The GTO is composed of a set of scientific themes each including several observational programmes. This comprehensive scientific effort aims at exploring the diversity in planetary systems through measurements at a signal-to-noise sufficient for constraining theoretical models. The ultimate goal of the mission is to allow for a better understanding of planet formation and evolution as well as of the prospects for finding planets suitable for harbouring life. 
The GTO themes described in this section have been assembled by the CHEOPS Science Team and participating CHEOPS Consortium Board members over three years preceding the launch. As such, it represents the diversity of scientific interests and expertise present in a group of over 40 scientists who met regularly 3 to 4 times a year. The GTO is technically structured in different science topics to provide easy reading and tracking of the whole CHEOPS science. Each science topic is called a theme and is made of a series of specific programmes. Each programme has its scientific objective, but may share targets with other programmes. The number of orbits allocated to each theme has \change{evolved over time to match the scientific needs of each programme} and is expected to \change{continue to change with time.  While the number of orbits allocated to different themes may shift, the total number remains bound to the 80\% allocated share of the GTO}. The following provides a short description of each theme.

\paragraph{Search for transits} The aim is to search for transits of planets discovered by other techniques, in particular among those detected by radial velocity measurements. Monitoring these systems around the predicted transit times of their planets offers a straightforward way of obtaining both mass and radius for a sample of super-Earths and Neptunes orbiting bright stars. At the heart of the original CHEOPS proposal, it was considered the only path forward before PLATO \cite{Rauer:2014} to find the nearest transiting planets, including rocky bodies within the habitable zone of their host star. Today, with NASA’s TESS mission in operation \cite{Ricker:2014}, the context has changed significantly leading to an optimisation of the target list with adjustments foreseen following TESS discoveries. At the time of writing, roughly 15\% of the GTO orbits are dedicated to this theme. 

\paragraph{Improve mass-radius relation} The mass-radius relation is the first step towards the characterisation of planetary bulk properties \cite{Zeng:2019}. The knowledge of the planetary composition, in turn, is a key element to constrain planet formation models, as it can be used to demonstrate, for example, transport of material in the proto-planetary disc \cite{Thiabaud:2015}. In this context, the structure of low-mass planets is the most relevant. The goal of this theme is to determine the mass-radius relation of planets, focusing on objects with a small radius (and/or low mass if this one is already known), rare objects (extreme in density, unusual radius, etc.), and planets in multi-planetary systems. This theme opens up fantastic opportunities for synergy with TESS as it provides follow-up options if needed. At the time of writing, roughly 25\% of the GTO orbits are dedicated to this theme.

\paragraph{Explore} This theme includes a set of programmes related to planet detection. They aim at (1) exploring the architecture of systems hosting small planets in relatively long-period orbits via transit timing variation (TTV); 
 (2) studying in details dust clumps in edge-on debris disks around young stars; (3) detecting new planetary systems around bright stars, focusing on both multi-planet systems and systems hosting hot Jupiters; (4) enlarging the parameter space of both planets and host stars, with particular emphasis on hot stars; (5) discerning planet migration scenarios using TTV; (6) searching for exo-Trojans. At the time of writing, roughly 15\% of the GTO orbits are dedicated to this theme.

\paragraph{Characterise atmospheres} {\change{High-precision broadband visible photometry of exoplanet occultation and phase-curves have proven particularly insightful for the understanding of atmospheric processes, especially when coupled with infrared measurements. This is especially true for hot Jupiters, to understand whether the exoplanet flux observed at visible wavelengths is due to thermal emission leaking into the shorter wavelengths or to reflected light due to high-altitude condensates or Rayleigh scattering. One example is the hot Jupiter Kepler-7b, which shows a prominent occultation and phase-curve signal in the Kepler data but no signal at all in the infrared, as measured by Spitzer at 3.6 and 4.5 microns \cite{Demory:2011,Heng:2013}. This points to Kepler-7b harbouring high-altitude clouds or hazes. Such inference would not have been possible without (broadband) observations at visible wavelengths. Other examples include a recent study \cite{Wong:2020} which combines phase-curves of multiple hot-Jupiter phase-curves observed by TESS and shows a possible trend between equilibrium temperature, day-to-nightside temperature contrast and recirculation efficiency. Gaidos et al. \cite{Gaidos:2017} even showed that combining TESS and CHEOPS broadband observations could achieve the distinction between thermal emission and reflected light for some targets, due to the differences in these facilities' wavelength-dependent sensitivity. At the time of writing, roughly 25\% of the GTO orbits are dedicated to this theme.}

\paragraph{Search for features} Planets close to their host star are tidally distorted into an ellipsoidal shape. Potentially, this effect is detectable in the transit light curve which would allow the measurement of the planet’s Love number providing further insight into the planet’s internal structure \cite{Akinsanmi:2019,Hellard:2019}. The main challenge is to separate ellipsoidal deformation from stellar limb darkening effects. Another potential visible dynamical effect is tidal dissipation resulting in a secular shrinking of the orbit. Measuring orbiting changes would allow us to obtain a direct measure of the quality factor $Q$ of a star \cite{Jackson:2009}. Unexpected features (asymmetry, bumps, etc.) observed in transit light curves could also lead to the detection of moons and rings. At the time of writing, roughly 5\% of the GTO orbits are dedicated to this theme.

\paragraph{Ancillary science} Programmes with relevance to the analysis of exoplanets data, to the interpretation of properties of exoplanetary systems, and particular questions in planetology have been grouped under the theme “Ancillary Science”. It includes stellar physics programs (the study of stellar micro-variability, the derivation of precise limb darkening laws as a function of stellar temperature) affecting the measurement of exoplanet parameters, as well questions such as the frequency of planets around evolved stars, or the detection of rings, jets, dense comas and even atmospheres around Centaurs/TNOs in the Solar System. This science is used as a filler programme using otherwise unattributed orbits.

\subsection{Community access to CHEOPS: the Guest Observers Programme}
\label{subsec:science_GO}
The GO programme is administered by ESA, and is open to the world-wide scientific community, regardless of nationality or country of employment and including members of the CHEOPS Consortium. The majority of the observing time is available through annual announcements of opportunity (AOs) soliciting observing proposals. Evaluation and assessment of the proposals, together with recommendations for the award of observing time, are made by an ESA-appointed CHEOPS Time Allocation Committee that works independently of the CHEOPS Mission Consortium. Proposals are selected based on scientific merit, taking into account the suitability of CHEOPS for the proposed observations: they can cover any science topic that can be shown to be both addressable by the performance capabilities of CHEOPS and compatible with the mission’s constraints. 

In order to allow new targets to be proposed by the Community at any time during the mission, up to 25\% of the open time (up to 5\% of the CHEOPS science observing \change{time}) will be allocated as discretionary time, as part of the Discretionary Programme. This will be overseen by ESA. Proposals for the  Discretionary Programme need to meet the merit criteria of those submitted for annual AOs, and in addition comprise a single target of high scientific interest that has either been discovered or declared to be of high scientific interest, since the time of the most recent annual call. 

The first AO for observations to be made in the first year in orbit came out in March 2019, with the announcement of successful proposals made in July 2019. The second AO is foreseen to come out in the fourth quarter of 2020. The Discretionary Programme \change{opened} shortly after the end of the In-orbit Commissioning campaign.

\section{Requirements and Estimated Performances}
\label{sec:requirements}
As outlined in section \ref{sec:science}, the main science objectives of CHEOPS rest on the ability of the spacecraft to perform high-precision photometric measurements of specific stars already known to host planets. Since CHEOPS is intrinsically a follow-up mission, it is essential that the spacecraft covers as large a fraction of the sky as possible so as to maximise the number of potential targets available. Provided the transit time is known, CHEOPS has no restriction on the orbital period of the planet it can observe. However, given the difficulty of detecting small mass planets at large distances, it was clear that the number of planets with orbital periods larger than 50 days available for follow-up would be small. From these different considerations, a number of requirements on the necessary performance of CHEOPS have been derived and used in the design of the mission. In particular, the photometric \change{precision} and sky coverage were the main drivers of the design. We review both of them below. 

\subsection{Photometric Precision and Stability}
\label{subsec:photometric_accuracy}
CHEOPS will observe different types of stars of different magnitudes. However, requirements on photometric precision and stability have been derived for magnitudes ranging from 6 to 12 in the V band. While stars brighter than magnitude 6 and fainter than 12 can be observed by CHEOPS, no precision nor stability requirements have been imposed on the measurements of these stars. The following are the top-level science requirements related to photometric precision and stability. 

\begin{description}
\item[\bf Bright stars (Science requirement 1.1):]
CHEOPS shall be able to detect Earth-size planets transiting G5 dwarf stars (stellar radius of \(0.9\,R_{\odot}\)) with V-band magnitudes in the range \(6\leq V\leq 9\,\)mag. Since the depth of such transits is 100 parts-per-million (ppm), this requires achieving a photometric precision of \SI{20} {ppm} (goal: \SI{10} {ppm}) in 6 hours of integration time (at least, signal-to-noise ratio of 5). This time corresponds to the transit duration of a planet with a revolution period of 50 days. 
\label{scireq:1.1}

\item[\bf Faint stars (Science requirement 1.2):]
CHEOPS shall be able to detect Neptune-size planets transiting K-type dwarf stars (stellar radius of \(0.7\,R_{\odot}\)) with V-band magnitudes as faint as V=12 mag (goal: V=13 mag) with a signal-to-noise ratio of 30. Such transits have depths of 2500 ppm and last for nearly 3 hours, for planets with a revolution period of 13 days. Hence, a photometric precision of 85 ppm is to be obtained in 3 hours of integration time. This time corresponds to the transit duration of a planet with a revolution period of 13 days.

\item[\bf Photometric stability (Science requirement 1.3):]
CHEOPS shall maintain its photometric precision for bright and faint stars during a visit (not counting interruptions), with a duration of up to 48 hours (including interruptions).

\end{description}

\subsection{Estimated Performances}
\label{subsec:performances}
To check whether the requirements above would be met, the overall performance of the instrument has been estimated in a semi-analytical way, considering results obtained from simulations and measurements with the aim of characterising the total noise for a set of benchmarks and comparing them with the science requirements (see \cite{Deline:2020} and \cite{Futyan:2020}). The sources of the different noise contributors have first been classified as being either astrophysical or instrumental in nature. The total noise budget results from the combination of all these contributors. The identified noise sources (see below) can be assumed to be independent of each other and thus added in quadrature. The total expected noise is then: 

\begin{equation}
\label{eq:noise}
N= \sqrt{\sum N_i^2}. 
\end{equation}

where $N_i$ represents a particular noise source as listed below:

\begin{enumerate}
    \item Astrophysical noise
    \begin{itemize}
        \item photon noise,
        \item zodiacal light, 
        \item cosmic rays,
        \item stray light parasitic illumination.
    \end{itemize}
    
        \item Instrumental noises:
    \begin{itemize}
        \item Point spread function in combination with the pointing jitter and the flat field uncertainties\footnote{\change{The pointing jitter requirement specifies that the instrument line of sight shall move less than 4 arc seconds RMS under nominal operations. In-flight measurements confirm that the jitter is much smaller than the requirement. See Sect.~\ref{subsec:spacecraft_design} for details.}},
        \item read-out noise (CCD plus analog chain random noise),
        \item dark current shot noise and dark current variation with temperature, 
        \item CCD gain and quantum efficiency stability,
        \item analog electronics stability,
        \item timing uncertainty,
        \item quantisation noise of the front-end electronics.
        
    \end{itemize}
\end{enumerate}
Note that the calculation of the total noise does not account for stellar intrinsic variations (flicker noise, stellar oscillations, etc.), contamination due to background stars, severe pixel defects, etc. \change{The stellar flicker noise, although not included in this calculation, is discussed in \cite{sulis:2020}}.
  
For a quantitative estimation of the noise budget, the stars are considered to be in the visual magnitude range between V = 6 and V = 12, and spectral types between G0 ($T_{eff} \sim 5900 \, K$) and M0 ($T_{eff} \sim 3800 \, K$).

\begin{figure}[htb]
\centering
  \includegraphics[angle=-90,width=0.95\textwidth]{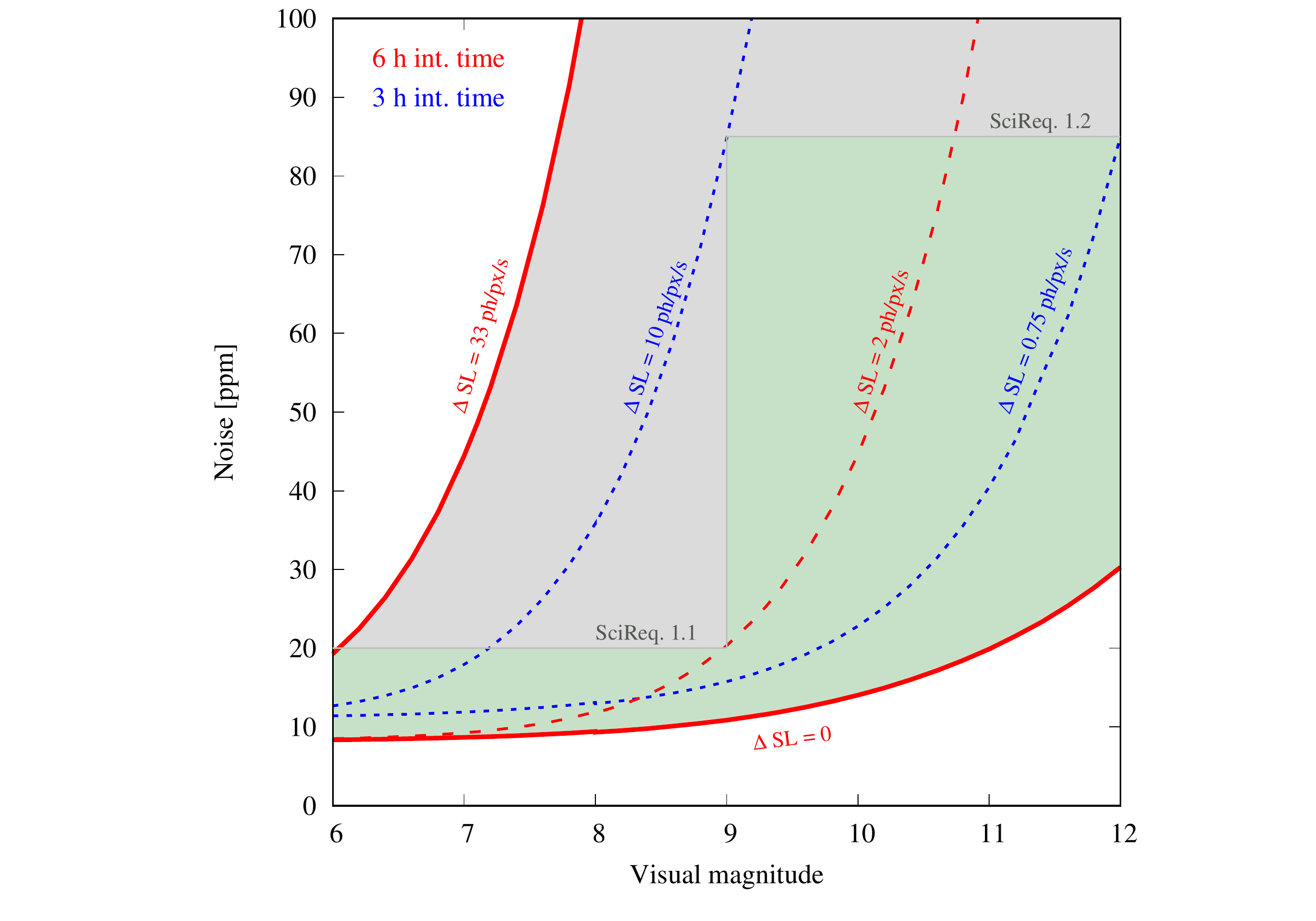}
\caption{Photometric noise as a function of the stellar magnitude. The solid red curve labelled $\Delta$\emph{SL} = 0 (no stray light contamination) represents the noise floor of CHEOPS (calculated for an M0 star). The dashed curves are examples of noise behaviour as a function of the stray light contamination variation along the satellite's orbit. The solid red curve labelled $\Delta$\emph{SL} = 33 ph/px/s shows the noise behaviour when the stray light contamination dominates the total noise for all magnitudes. The green region defines the conditions that satisfy the performance science requirements. }
\label{fig:noise}       
\end{figure}

Figure~\ref{fig:noise} shows the results of the expected total noise after 3 hours (blue curves) or 6 hours (red curves) of integration. These estimations can be compared to the actual measurements obtained during in-orbit commissioning (Sect.~\ref{subsec:photometric_precision}). Among the different noise sources, the parasitic Earth stray light,  i.e. the sunlight reflected from the Earth's surface, is the one whose contribution can vary the most because it depends upon the pointing direction and on the Sun's position with respect to the Earth at the time of observation. The Earth stray light can vary, in one orbit, from zero (the telescope is flying over a dark region of the Earth surface) to thousands of photons per pixel per second (the telescope is pointing close to the illuminated Earth limb). Even if the post-processing of the images allows for correction of more than 99.5\% of the background contamination, the variation along one orbit of the stray light flux can increase dramatically the total noise. 

In Figure~\ref{fig:noise}, the noise floor is shown by the red solid curve labelled $\Delta$\emph{SL} = 0. This curve was calculated for an M0 star considering 6 hours of uninterrupted observations and no stray light contamination during the whole observation. For bright targets, the instrumental noise dominates the total noise while for the fainter ones the photon noise is the main contributor. Therefore, the curve $\Delta$\emph{SL} = 0, represents the minimum noise expected for a CHEOPS standard target (note that for earlier spectral types the photon noise increases).  The red dashed curve also accounts for 6 hours of uninterrupted observations but for a G0 star and a maximum stray light variation of 2 photons/pixel/second along the orbit. The impact of adding this amount of stray light variation to the noise budget is that its associated noise starts to dominate the noise budget for stars fainter than magnitude 9. For faint stars to be observable with high signal to noise ratio, the stray light flux variation has to be really low, as shown in the rightmost dashed blue curve (3 hours of integration time, M0 star and a maximum stray light variation of 0.75 photons/pixel/second). In this case, the photon and instrument noises shape the noise budget for almost the whole magnitude range except in the faint end, where the stray light variation that dominates, contributing with 70 ppm to the total noise for a magnitude 12 star. The leftmost dashed blue curve shows an example (3 hours integration time, G0 star, 10 photons/pixel/second stay light variation) where the stray light contribution to the noise budget dominates over the other noise sources in all cases.   Finally, the solid red curve sets the noise ceiling for 6 hours integration time (G0 star,  $\Delta$\emph{SL} = 33 ph/px/s ) where all other noise contributors are negligible compared to the stray light noise. It is therefore evident that small variations in the stray light level along the orbit can increase significantly the total noise. 
  
Given the impact that stray light variations can have on the noise budget, the stray light contribution should be accounted for when scheduling an observation. Special care has been taken to guarantee that the schedule solver avoids planning an observation when high stray light variations are expected to impact the images (i.e. observations are scheduled to fall within the green region of the diagram).

\subsection{Orbit and Sky Coverage}
\label{subsec:sky_coverage}
Ultimately, the number of targets that CHEOPS will be able to follow depends upon the fraction of the sky that can be observed while meeting the photometric precision described above (see section \ref{subsec:photometric_accuracy}). At the time the mission was proposed, most planets orbiting stars with magnitudes in the V = 6-12 range were detected using radial velocity techniques, the huge number of planets discovered by the Kepler spacecraft being significantly fainter. Another source for targets was expected to be ground-based transit detections, especially from Next Generation Transit Survey (NGTS) located at Paranal Observatory in the Southern hemisphere. With these considerations in mind, the following are the two top-level requirements defined for sky coverage. 

\begin{description}
\item[\bf Planets discovered by radial velocity measurements:]
50\% of the whole sky shall be accessible for 50 (goal: 60) cumulative (goal: consecutive) days per year and per target with time spent on-target and integrating the target flux longer than 50\% of the spacecraft orbit duration.

\item[\bf Planets discovered by ground-based transit measurements:]
25\% of the whole sky, with 2/3 in the southern hemisphere, shall be accessible for 13 days (cumulative; goal: 15 days) per year and per target, with time spent on-target and integrating the target flux longer than 80\% of the spacecraft orbit duration.
\end{description}

To comply with this requirement on sky coverage, a Sun-synchronous, dawn-dusk orbit with a local time at the ascending node of 6am was chosen with a possible altitude ranging between 650 and 800 km. Later, the choice converged on a final altitude of 700 km (actual orbit: 7078.848 km mean semi-major axis, alt 700.713 km, with an orbital period of 98.725 minutes). In fact, this orbit represents a trade-off between meeting the sky coverage requirement, minimising stray light by having the line of sight as much as possible over the dark side of the Earth and reducing the radiation environment. Finally, the availability of a suitable launch opportunity and compliance with space debris mitigation regulations also played a role in orbit definition.

\begin{figure}[htb]
\centering
  \includegraphics[width=0.95\textwidth]{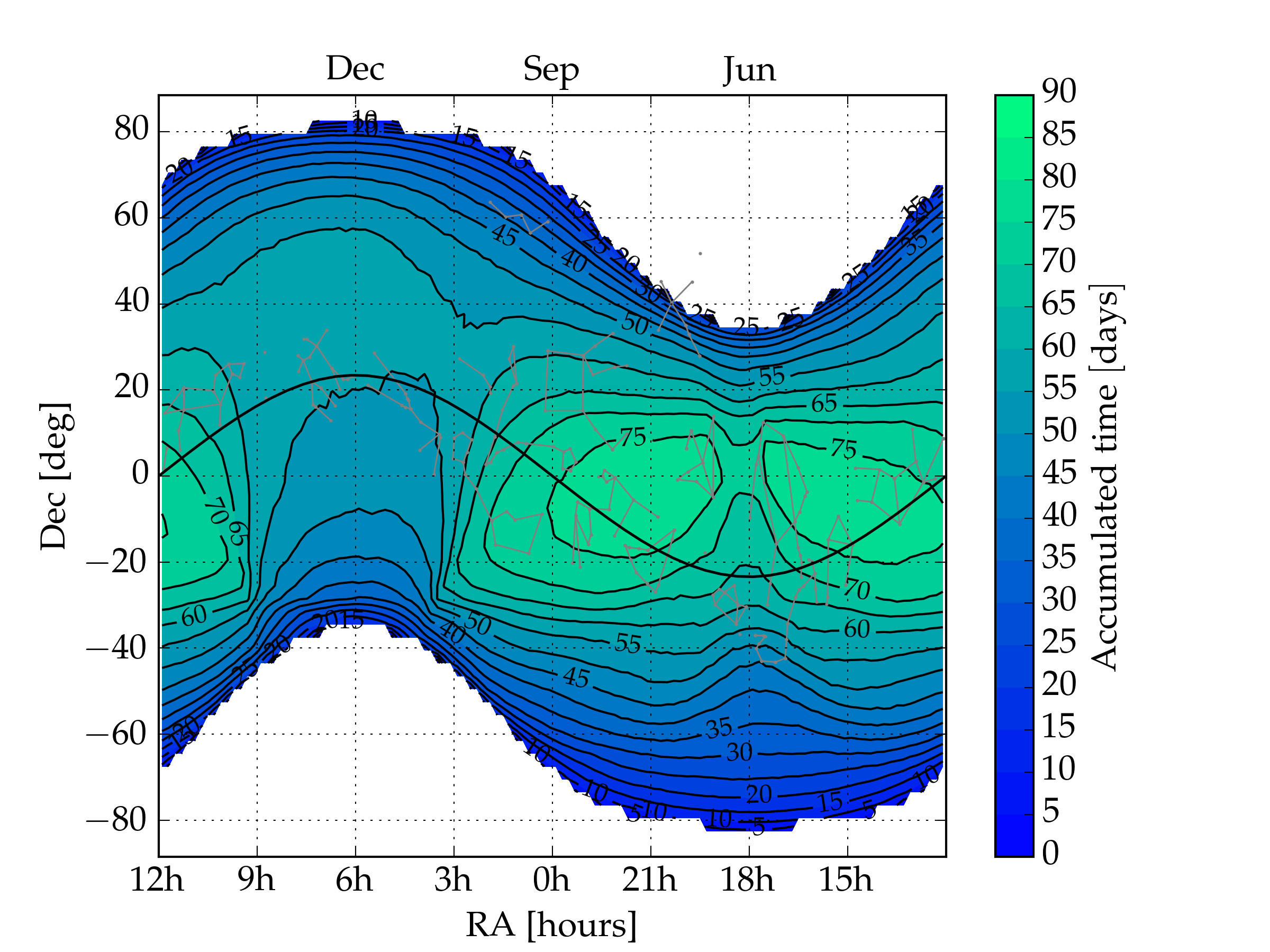}
\caption{CHEOPS visibility map. The color code shows the accumulated time over one year for each possible pointing direction. Note that orbits with more than 50\% interruptions were discarded for computing this map. December, September and June are marked as a reference at the top of the plot to indicate when a certain region in the sky is observable. Zodiacal constellations are over-plotted in grey for reference around the ecliptic (thick black line), together with a few other useful constellations. (In particular, the Galactic plane can be visualised as crossing Sagittarius and following the neck of Cygnus in the Summer sky, while crossing the ecliptic close to Taurus in the Winter sky). Note that the Kepler fields are essentially out of reach for CHEOPS}
\label{fig:sky} 
\end{figure}

While the selected orbit also minimises the Earth occultations of the sky, another restriction associated with the strict control of the temperature of the CCD further limits the sky coverage. The CCD is cooled passively by radiating to deep space any excess heat. To allow for this, the Sun must at all times remain outside a cone of half-angle of \ang{120} centred on the line of sight of the telescope. This requirement further limits the pointing of the telescope making it impossible to observe close to the ecliptic poles. 

Finally, during the observation of a target, the flow of usable measurements (images) can be interrupted due to:
\begin{itemize}
    \item the target being occulted by the Earth during part of the satellite's orbit;
    \item the stray light contamination is too high;
    \item cosmic ray hits during the passage through the South Atlantic Anomaly being too high. 
\end{itemize}

These interruptions will appear as gaps in the light curve. To conserve bandwidth most of the data taken during these times will not be down-linked to the ground. 

With all this considered, the sky visible to CHEOPS can be calculated.  Figure \ref{fig:sky} shows a map of the annual sky coverage. 

\section{Project Implementation Approach}
\label{sec:implementation}
The CHEOPS mission is a partnership between ESA's Science Programme and Switzerland, with important contributions from Austria, Belgium, France, Germany, Hungary, Italy, Portugal, Spain, Sweden, and the United Kingdom. All these ESA member states constitute the CHEOPS Consortium. Their contributions to the mission in terms of hardware or software are estimated to match the cost-capped ESA budget of 50 MEUR foreseen for small missions. Hence, the total available CHEOPS budget is slightly over 100 MEUR. This amount includes designing, building, launching, and operating the CHEOPS satellite for a nominal period of 3.5 years. 

In response to the CHEOPS development challenges, driven by the limited (and cost-capped) budget and by the short development time, a number of measures were adopted in setting up the project organisation and in defining the implementation approach (see also Sect.~\ref{sec:mission_SC_design}). The key measures, for different project areas, are summarised below.
\begin{enumerate}
    \item Project organisation: small size teams were deployed to maintain close coordination as well as enable a faster decision process. 
    \item Technology readiness:  the mission had to be compatible with the re-use of an existing “off-the-shelf” platform and had to include a payload based on available technologies (\change{Technology Readiness Level} $>$ 5-6 in ISO scale), ruling out complex development activities and focusing on implementation and flight qualification aspects.  
    \item Realism \& stability of requirements: both the accelerated development schedule and the mission cost ceiling required realistic and stable requirements from the very start of the project. The requirements were consolidated by the System Requirements Review (SRR) and, later in the project, significant efforts were made to avoid modifying or adding requirements. 
    \item Industrial implementation approach: as described in \cite{Rando:2018}, and in contrast to the M- and L-class missions, a single Invitation to Tender was issued for CHEOPS, covering both a parallel competitive study phase (A/B1) and the implementation phase (B2/C/D/E1, including responsibility for Launch and Early Operation Phase - LEOP and In-Orbit Commissioning - IOC). The prime contractor was selected shortly after the SRR. 
    \item Early mission concept definition: the industrial procurement approach described above (single tender with a ceiling price covering also the implementation phase) is only feasible when the mission concept and the space segment requirements and early design are mature enough to rule out major changes in later phases. A concurrent engineering approach (in the form of a phase 0/A study performed at the Concurrent Design Facility at ESTEC) was applied in order to achieve the required mission concept and S/C requirements maturity in less than 6 months from proposal selection.
    \item Definition of interfaces: the early definition of clear and stable interfaces (instrument-platform, spacecraft-ground, spacecraft-launcher) was requested from all parties as an essential pre-condition to ensure that the different teams could work in parallel and meet their challenging development schedules. Frequent interface technical meetings have been organised with ESA, as mission architect, playing an important coordination role.
    \item  Review cycle: the CHEOPS project has followed the standard ESA review cycle. However, in order to maintain compatibility with the stringent schedule constraints, the duration of the reviews was compressed compared to larger missions, adapting the number of panels and reviewers to streamline the process.
\end{enumerate}

This implementation approach proved effective: it has allowed the selection of the prime contractor and to start the implementation phase (B2/C/D/E1) in April 2014, after only 1.5 years from mission proposal selection, to complete the system Critical Design Review (CDR) in May 2016 (2 years after) and to complete the satellite level test campaign in December 2018 (about 6 years after the proposal selection). Figure \ref{fig:del_milestones} summarizes the key project development milestones. It should also be noted that the newly designed CHEOPS payload was delivered to the prime contractor in April 2018, in approximately 5 years and a half from mission selection, and in less than 4 years from instrument level Preliminary Design Review (PDR). The satellite was declared flight-ready in February 2019, less than 3 years after CDR \cite{Rando:2019}. The successful launch took place on December 18, 2019, nine months after the Qualification \& Acceptance Review (QAR). CHEOPS being launched as a secondary passenger, this launch date was entirely driven by the readiness of the prime passenger. 

\begin{figure}[htb]
  \includegraphics[width=1.0\textwidth]{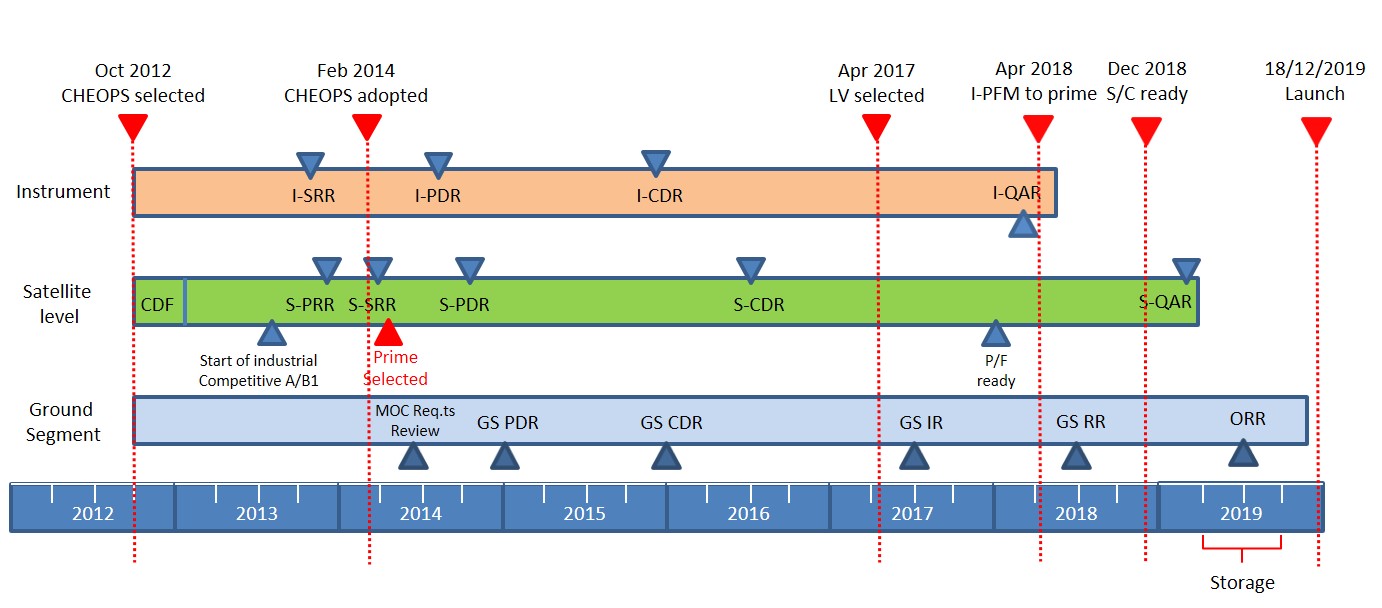}
\caption{Summary of the CHEOPS development milestones.}
\label{fig:del_milestones}       
\end{figure}

\section{Mission and Spacecraft Design}
\label{sec:mission_SC_design}
\subsection{Mission Design}
\label{subsec:mission_design}
The CHEOPS mission is designed to operate from a dawn-dusk, Sun-synchronous orbit, at an altitude of 700 km. As mentioned above, this orbit was selected during the assessment phase to maximise sky accessibility, to minimise stray-light, and to reduce the radiation environment while ensuring the largest possible number of shared launch opportunities as well as compatibility with existing platforms, qualified for \change{Low Earth Orbit (LEO)}. 

The nominal mission design lifetime is 3.5 years with a possible mission extension to 5 years. No consumables are used for nominal operations. Thus, the dominant factors limiting the lifetime of the mission are considered to be linked to radiation damage to the detector as well as overall component degradation due to exposure to cosmic radiation and ageing in general. The platform includes a compact, mono-propellant propulsion module, required to perform an initial launcher dispersion manoeuvre, to enable collision avoidance manoeuvres and to comply with the space debris mitigation regulations, re-entering the S/C within 25 years from the end of operations. The spacecraft design is described in the following section. 

The instrument-to-platform interfaces are based on isostatic mounting of the instrument Baffle Cover Assembly (BCA) and of the Optical Telescope Assembly (OTA), both accommodated on the top panel of the platform; the instrument is thermally decoupled from the rest of the spacecraft (with the exception of two instrument electronic units installed inside the platform); the optical heads of two star-trackers are mounted directly on the OTA structure (to minimise misalignment effects induced by thermo-elastic distortion).  

The spacecraft configuration was driven by the installation of the instrument on top of the platform, behind a fixed Sun-shield, also supporting the Solar Arrays, and with compact dimensions, so as to fit within different launch vehicle adapters (in particular under ASAP-S on Soyuz and under VESPA on VEGA) and to guarantee the capability to point the line of sight within a half-cone of 60 deg centred around the anti-Sun direction. It should be noted that the Soyuz launcher was selected after the satellite CDR.

\subsection{Spacecraft Design}
\label{subsec:spacecraft_design}
The satellite design is based on the use of the AS-250 platform, an Airbus Defence \& Space product line designed for small and medium-size missions operating in LEO. Airbus Spain is the  prime contractor \cite{Asquier:2018}. The spacecraft configuration (see Figure \ref{fig:CHEOPS_config}) is characterised by a compact platform body, with a hexagonal-prismatic shape and body-mounted solar arrays, which also maintain the instrument and its radiators in the shade for all nominal pointing directions within a half-cone of 60 deg centred around the anti-Sun direction.

The configuration has been optimized to be compatible with both the ASAP-S and VESPA adapters for a shared launch respectively on Soyuz or VEGA: the spacecraft is just above \SI{1.5}{m} tall and has a footprint remaining within a circle of \SI{1.6} {m} in radius. The total wet-mass is close to \SI{275} {kg}. Such dimensions and mass are in line with the key requirement of maintaining compatibility with different small and medium-size launcher vehicles as an auxiliary or co-passenger. In fact, the spacecraft was designed to be compatible with launch environment requirements enveloping different launchers. This approach proved pivotal for the CHEOPS project, considering that the launcher selection was finalised in 2017 (Soyuz, under ASAP-S). 

The hexagonal-prismatic platform features vertical beams, corner joints, and lateral panels which can be opened to facilitate the equipment integration (Figure \ref{fig:CHEOPS_config}, right side). The three lateral panels in the anti-Sun direction are equipped with radiators procured from Iberespacio (ES). 

\begin{figure*}[htb]
  \includegraphics[width=0.97\textwidth]{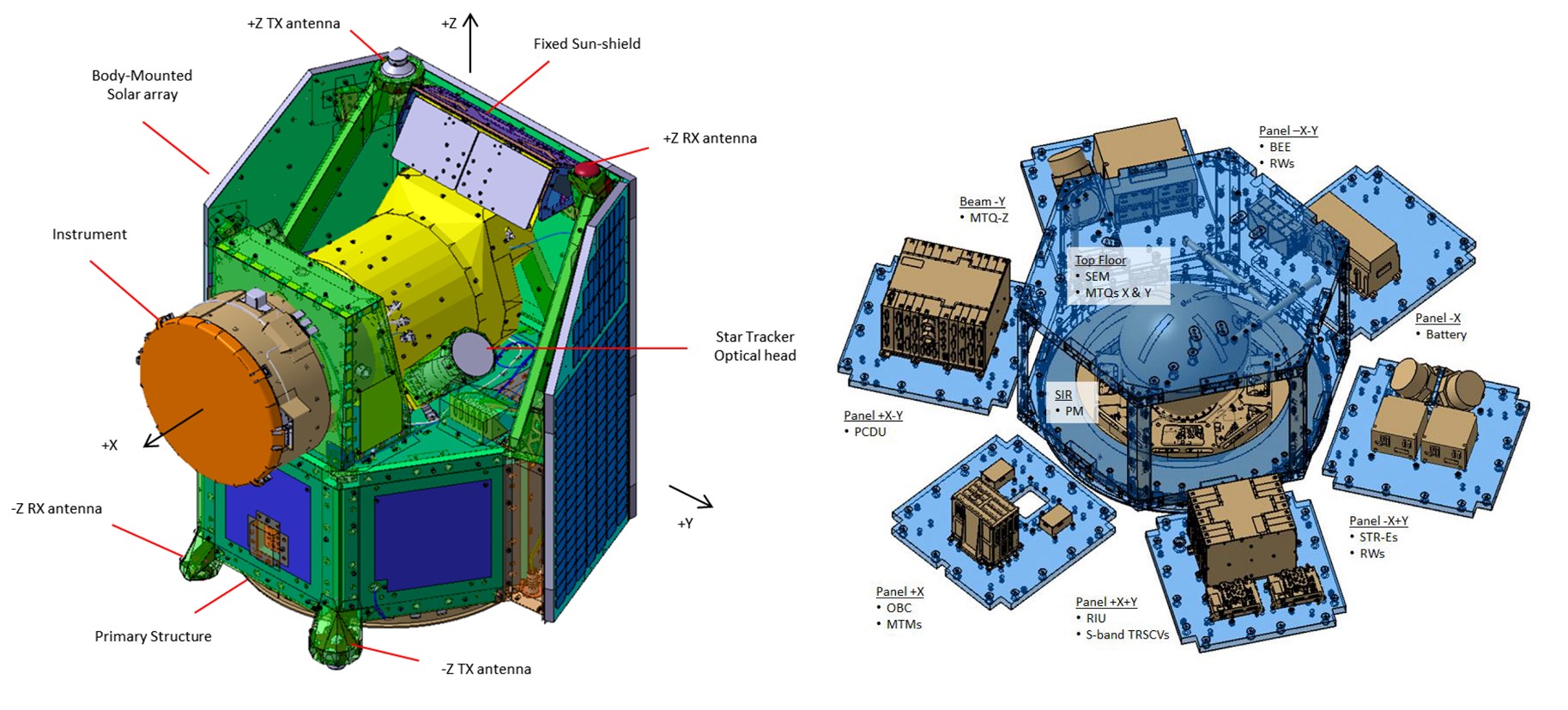}
  \caption{CHEOPS spacecraft configuration and units accommodation in the platform (courtesy of ADS-Spain).}
\label{fig:CHEOPS_config}       
\end{figure*}

The key avionic units (On-Board Computer, Remote Interface Unit, Power Control \& Distribution Unit) and On-Board Control Software (adapted to the mission's specific needs) are inherited from the AS-250 product line. Part of the Attitude and Orbit Control System (AOCS) equipment also belongs to the AS-250 line, with the exception of the reaction wheels (provided by MSCI, Canada), the absence of GPS, and Coarse Sun Sensor, and the down-scaling of the magneto-torquers (Zarm, DE). The two Hydra star trackers have been provided by Sodern (FR) and have been installed on the instrument optical telescope assembly to minimise misalignment to the instrument line of sight induced by thermo-elastic distortion. 

Two star trackers monitor the position of the stars in the sky to allow for orientation of the spacecraft. For improved performance, their optical heads have been directly mounted on the payload structure. Furthermore, the \change{Attitude and Orbital Control System (AOCS)} uses the instrument as a fine guidance sensor. The measured difference between the real centroid position of the target star and the expected position is fed back to the AOCS to improve the pointing stability in particular for longer observations. The requirement on the pointing precision in this mode hacentreds been set \ang{;;4} RMS over a 48h observing period. In practice, during commissioning a pointing accuracy of order \ang{;;1} could be achieved. During observations the instrument line of sight is aligned to the apparent position of the targeted star. Rotation of the platform around the line of sight was implemented so that the instrument radiators, needed to maintain thermal balance of the instrument, are always directed towards deep space thereby receiving as little as possible infrared and reflected light from the Earth. Finally, for safety reasons and to minimise stray light, the platform is allowed to point the instrument in any directions lying within a half-cone of 60 deg centred around the anti-Sun pointing direction.

The electrical power is provided by 3 solar arrays, two lateral panels, and a central one. The total geometric area is about 2.5 square meters. The Photo-Voltaic Assembly (Leonardo, IT) is based on 3G30 solar cells, which are installed on 3 sandwich panels with a Carbon Fiber Reinforced Plastic skin; the assembly is sized for an average power of just below 200 W (in nominal mode) and takes into account ageing and degradation effects over the mission lifetime.

The CHEOPS spacecraft includes a compact, mono-propellant Propulsion Module (PM) from Arianespace Group (DE), inherited from the Myriade Evolution design and including a Hydrazine tank with a capacity of 30 litres from Rafael (IS), 4 small 1N thrusters, 1 pressure transducer and 2 pyro-valves (one of them for passivation at end of line). The PM has been assembled and tested as a separate sub-system before integration on the satellite. Its mechanical interface, directly on the Satellite Interface Ring, has been designed to ensure maximum modes decoupling between the PM and the rest of the S/C. The propellant tank has been sized for providing a total delta-V in excess of the total of 110 m/s required to perform: a) launcher dispersion correction manoeuvre, b) collision avoidance manoeuvres and c) final de-orbiting at the end of the operational phase.

The telecommunication sub-system is based on a redundant S-band transceiver (COM DEV, UK) \cite{Rando:2018}, \cite{Asquier:2018}, with two sets of RX and TX antennas provided by RYMSA (ES) and located respectively at the +Z and –Z  end of the S/C (see Figure~\ref{fig:CHEOPS_config}).

\section{The CHEOPS Payload}
\label{sec:CHEOPS_PAYLOAD}
The CHEOPS payload consists of a single instrument: A high performing photometer measuring light variations of target stars to ultra-high precision \cite{Beck:2017}. The photometer is operating in the visible and near-infrared range (\SIrange{0.4}{1.1}{\micro \metre}) using a back-illuminated CCD detector run in Advanced Inverted Mode Operation (AIMO). The instrument is composed of four different units. The telescope together with the baffle is mounted on top of the platform while the Sensor Electronic Module (SEM) and the Back End Electronics (BEE) are hosted inside the platform body. 

The telescope design is based on an on-axis Ritchey-Chr\'{e}tien optical configuration with a \SI{320}{mm} diameter primary mirror. Considering the central obscuration of the primary mirror, the effective collecting area of the system is \SI{76793}{mm^2}.

\begin{figure}[htb]
  \includegraphics[width=1.0\textwidth]{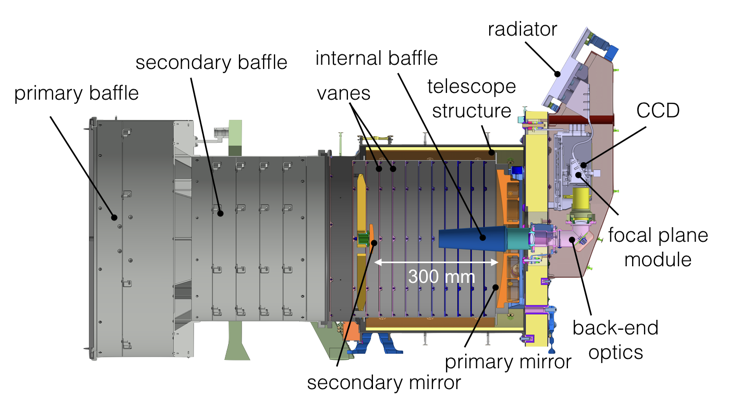}
\caption{CAD/CAM view of the OTA and BCA assemblies as mounted on the spacecraft. The primary and secondary baffle constitute the BCA. It is a separate unit mounted on the spacecraft independently of the OTA. The remaining items resemble the OTA.}
\label{fig:OTABCA}       
\end{figure}

\subsection{Optical Telescope Assembly (OTA)}
\label{OTA}

The OTA hosts the optics as well as the detector and the read-out electronics of the instrument. The optical configuration consists of a Ritchey-Chr\'etien telescope and a Back End Optics (BEO) to re-image the telescope focal plane on the detector and to provide an intermediate pupil where a pupil mask is placed for stray-light rejection. In order to keep the baffling system long enough in front of the telescope a very fast (almost F/1) primary mirror fed the secondary making an intermediate F/5 telescope focal plane. The BEO enlarges the final effective focal length to about F/8.38 or an effective focal length of \SI{2681}{mm} at the nominal wavelength of \SI{750}{nm}.

Because of the short focal length of the primary mirror special care has been devoted to procedures for the Assembly, Integration and Verification (AIV) of the Ritchey-Chr\'etien configuration, because of the high sensitivity of the secondary mirror alignment. This has been accomplished trough the realization of a demonstration model equivalent from the optical point of view to the flight one. The AIV procedures have been tested and refined on the model \cite{Bergomi:2014,Bergomi:2016} and later incorporated into the actual flight telescope \cite{Bergomi:2018}.

In the development phase an holographic device able to smear out uniformly on a squared region the light of the observed star has been considered \cite{Magrin:2014} but it has been discarded because it was non compatible with the timescale of the project and the technological readiness level of such an approach.

The \ang{0.32;;} field of view is translating into a \SI{}{\ang{;;1} \per px} plate scale on the detector (\SI{13}{\micro m} pixel \(1k\times 1k\), AIMO). Figure~\ref{fig:OTABCA} illustrates the OTA/BCA assembly CAM/CAD cut view. The TEL group with the primary and secondary mirror, the BEO, and the focal plane module (FPM) are indicated. Figure \ref{fig:FM} shows the flight telescope with its mirror installed (right panel) and the fully assembled and integrated instrument just prior to its delivery in the CHEOPS Laboratory at the University of Bern (left panel). Finally, Figure \ref{fig:SC_accom} shows the instrument after integration on the platform at Airbus Defense and Space in Madrid. 

\begin{figure}[htb]
  \includegraphics[width=0.55\textwidth]{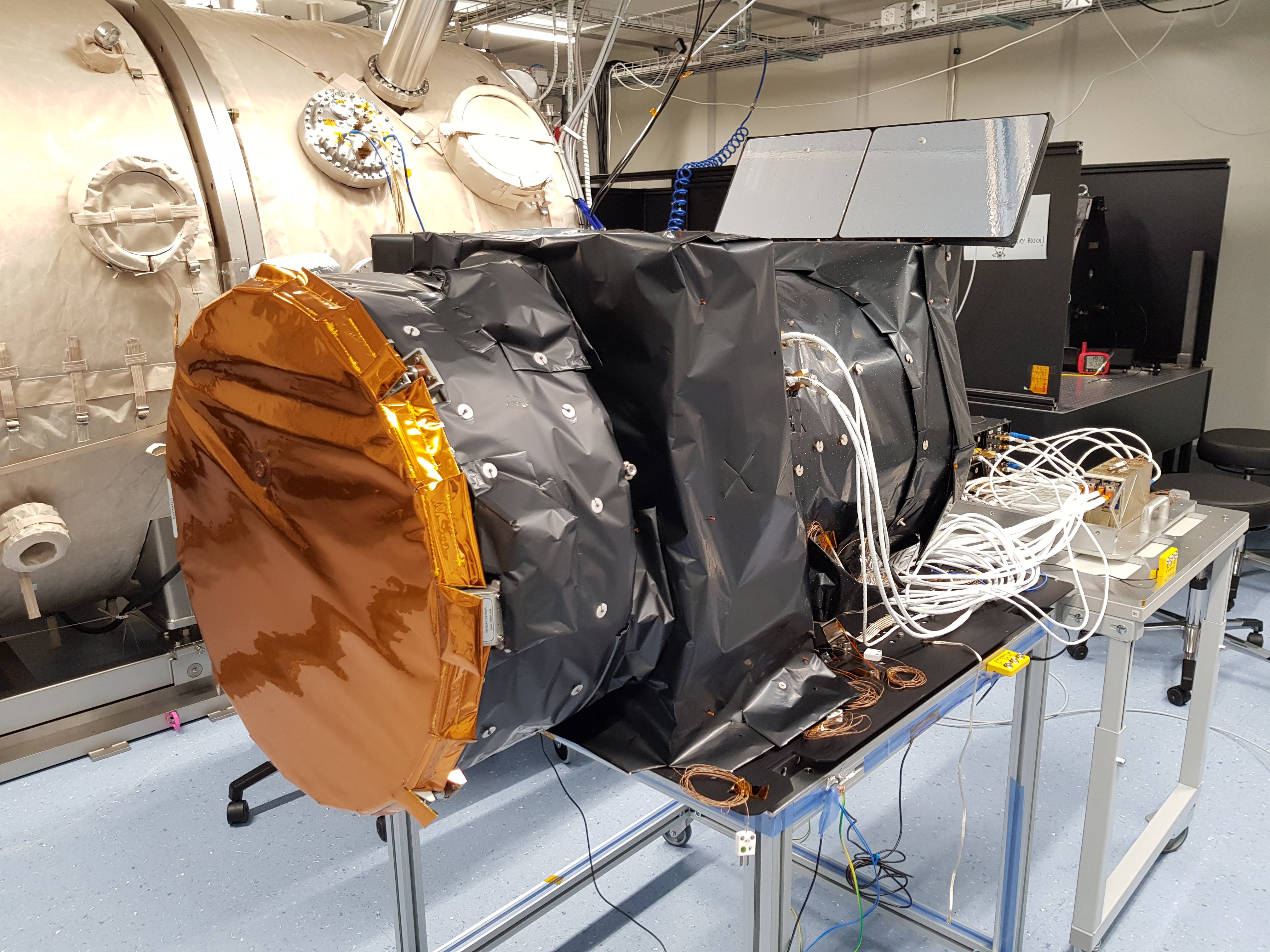}
  \includegraphics[width=0.45\textwidth]{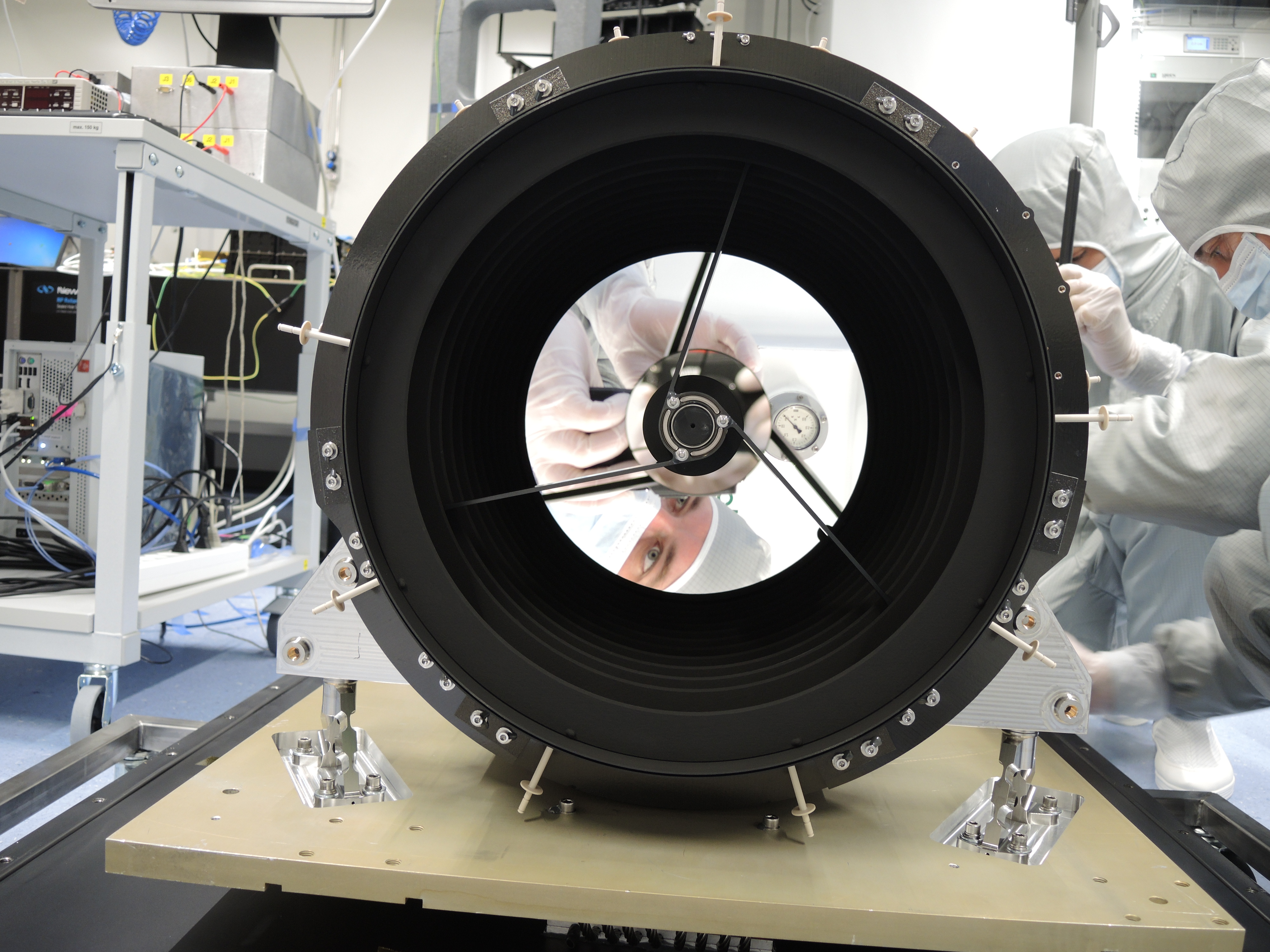}
\caption{Left: The CHEOPS payload fully assembled before delivery. Right: The telescope with mirrors installed and cover open before full OTA assembly.}
\label{fig:FM}  
\end{figure}

\begin{figure}[htb]
\centering
  \includegraphics[width=1.0\textwidth]{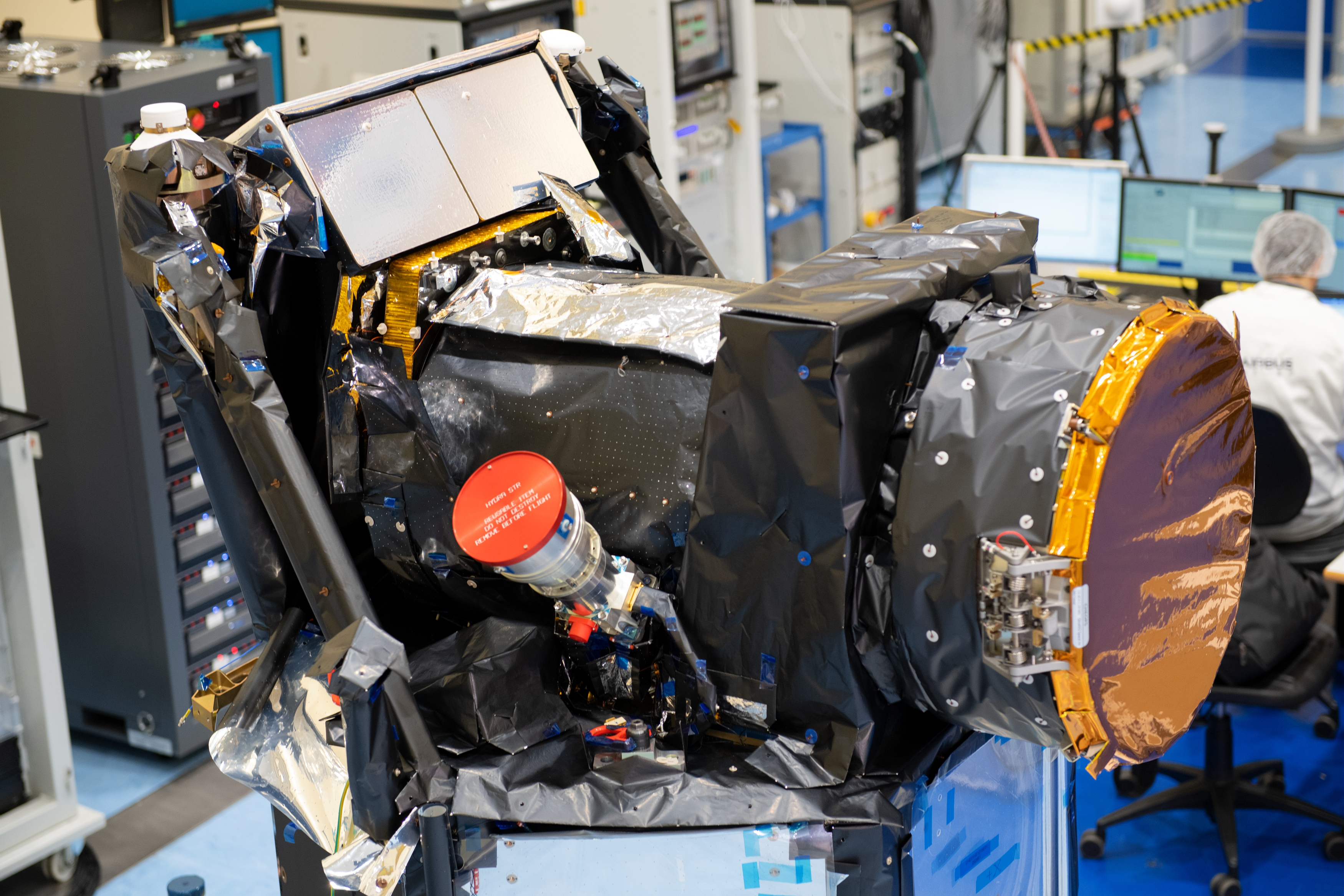}
\caption{Picture of the payload after integration on the platform with the star tracker optical heads (red covers). Image credits Airbus DS Spain.}
\label{fig:SC_accom}       
\end{figure}

The focal plane module hosts the CCD and the readout analogue and digital electronics. The detector is a single frame-transfer back-side illuminated E2V AIMO CCD47-20. The image section of the CCD has an area of $ 1024 \times 1024$ pixels, while a full-frame image including the covered margins is 1076 columns and 1033 rows as represented in Figure~\ref{fig:ccd}. The sides and top margins of the CCD consist of 16 covered columns of pixels (called dark pixels) and 8 columns of ``blank'' pixels (they are not real pixels but electronic registers) at each side. On the side opposite to the output amplifier, there are 4 virtual columns of overscan pixels (left side margin in the figure). These pixels are not physical pixels of the CCD but generated by 4 additional shift-read cycles. 

Due to the restricted telemetry capabilities of CHEOPS, it is not possible to download all full-frame images. Instead, window images of $200 \times 200$ pixels centred on the target star plus the corresponding covered margins are cropped from the full-frame image. Note that to further save bandwidth, the on-board software will crop the corners of the window images so that circular window images (of diameter 200 pixels) are actually sent to the ground. Figure~\ref{fig:ccd} shows a schematic of the CCD, full-frame image, and window image.

\begin{figure}[htbp]
\centering
 \includegraphics[width=0.75\textwidth]{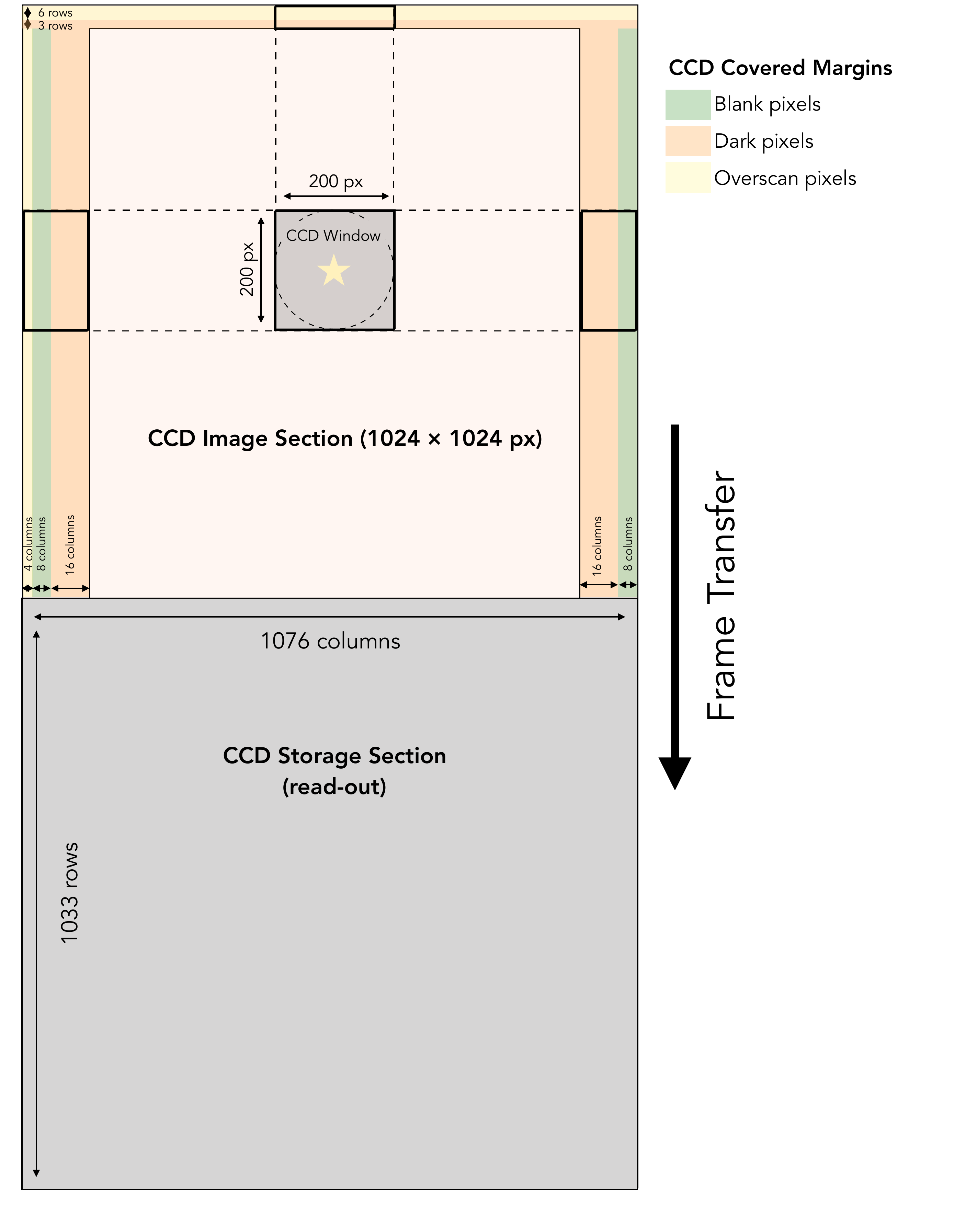}
\caption{Schematic view of the CCD elements.}
\label{fig:ccd}
\end{figure}

The nominal operating temperature of the CCD was initially set at \SI{233}{\K} and its temperature stabilised within \SI{10}{\milli \K} while the read-out electronics is stabilised to \SI{50}{\milli \K}. During in-orbit commissioning it was decided to lower this operating temperature to \SI{228}{\K} to reduce the number of hot pixels (see Sect.~\ref{ioc}). The excess heat generated by the focal plane module is radiated to deep space using two passive radiators located on top of the OTA. The requirement for temperature stabilisation comes from the need for having a very stable and low noise CCD read-out. Low operating temperatures minimise the read-out and dark noise in the system while temperature stability minimise dark noise variation and especially the CCD gain variability, changes in quantum efficiency, as well as the analogue electronics overall (e.g. CCD BIAS voltage stability).

To improve photometric performances by mitigating the effect of pointing jitter, avoiding saturation for bright targets, and to mitigate imperfect flat-field correction, CHEOPS has been purposely defocused. The size of the point-spread function (PSF), that is the radius of the circle enclosing 90\% of the energy received on the CCD from the target star, has been estimated during the design phase to be within \SIrange{12}{15}{px}. The actual value as measured during in-orbit commissioning is \SI{16}{px} (see Sect.~\ref{subsec:psf}), a value reasonably close to the expected value. 

Note that de-focusing the telescope has also some drawbacks. During construction, due to the divergence of the beam once out of focus, it made aligning the focal plane to the desired PSF size more challenging. Precise photometry is also more difficult to obtain in very crowded fields and for very faint stars for which the light is spread over many pixels. 

\subsection{Baffle and Cover Assembly (BCA)}
The BCA is key to the stray light mitigation strategy adopted for the payload. The baffle design has been adapted from the CoRoT baffle but scaled down in size and adjusted for the on-axis design of the CHEOPS telescope. The baffle consists of a cylindrical aluminum tube with a number of circular vanes and black coating for stray-light rejection. The baffle is terminated by a cover assembly which protects the optics from contamination during assembly, integration, and tests as well as during launch and early operations phase. The cover release is a one-shot mechanism based on a spring-loaded hinge and a launch lock mechanism. The launch lock makes use of a Frangibolt actuator design. Figure~\ref{fig:FM} shows an image of the instrument. The view shows the fully assembled BCA/OTA and a telescope view with cover open after mirror installation. On the left side of the BCA the hinge can be seen, Figure~\ref{fig:SC_accom}. The Frangibolt is enclosed on the right hand side.

The BCA and OTA are separately mechanically mounted on the top deck of the platform. The BCA to OTA interface is sealed by a Vetronite seal in order to close off the complete optical cavity. The baffling system of the instrument, including the OTA baffling is designed to be able to reject  stray light impinging at an angle greater than \ang{35} with respect to the optical axis by several orders of magnitude. The level of stray light suppression required by the baffling system is in the range of \numrange{e-10}{e-12} depending upon the incidence angle. To meet this stray light rejection requirement, cleanliness is of key importance as dust will scatter light into the optical path. Therefore all the assembly, integration, and tests were performed in a class 100 clean room. Target optics cleanliness in orbit is 300 ppm. We note that the desired stray-light rejection factor was evaluated by simulations only as measurements were not feasible given the magnitude of the rejection factor.

\subsection{Sensor Electronics Module (SEM)}
\label{subsec:SEM}
The SEM is located inside the platform and controls the focal plane module. It is physically decoupled from the FPM to minimise the heat load on the telescope assembly. The SEM is hosting the Sensor Control Unit and a Power Conditioning Unit. It interfaces the Back End Electronics (BEE) and is commanded from it. The functions and characteristics of the SEM can be summarised as follows [5]:
\begin{itemize}
    \item Digital control electronics (FPGA including a LEON processing unit, RAM, EEPROM,
PROM) for clocking the CCD \change{and Front End Electronics}, reading
out the CCD and processing of data.
    \item Generate clocking modes for different CCD
read-out modes.
    \item Perform thermal control of the CCD and \change{Front End Electronics}
    \item Voltage conditioning towards the FPM
    \item Windowing of CCD image getting a 200x200
star window and overscan windows
    \item Time stamping of CCD and \change{Housekeeping} data.
    \item Sending reference star images to the BEE for
AOCS processing/interface test.
    \item Analog Housekeeping acquisition by a
separate \change{Analog-Digital Converter}.
\end{itemize}

For a high precision photometer, the CCD gain (the ratio between the number of electrons per pixel and the number of counts per pixel) needs to remain extremely stable. As part of the flight model CCD selection process and later the instrument calibration, high-precision color-dependent flat fields, the characteristics of the detection chain, gain, full-well capacity, read-out noise, have been measured in the laboratory at the Universities of Geneva and Bern \cite{Wildi:2017,Chazelas:2019}. Also, the instrument distortion has been characterised as well and the variation of the point-spread function across the field explored \cite{Chazelas:2019}.  Figure~\ref{fig:CCD_cryo} shows the CCD mounted inside the cryostat used for the flight unit characterisation.

\begin{figure}[htb]
\centering 
  \includegraphics[width=0.8\textwidth]{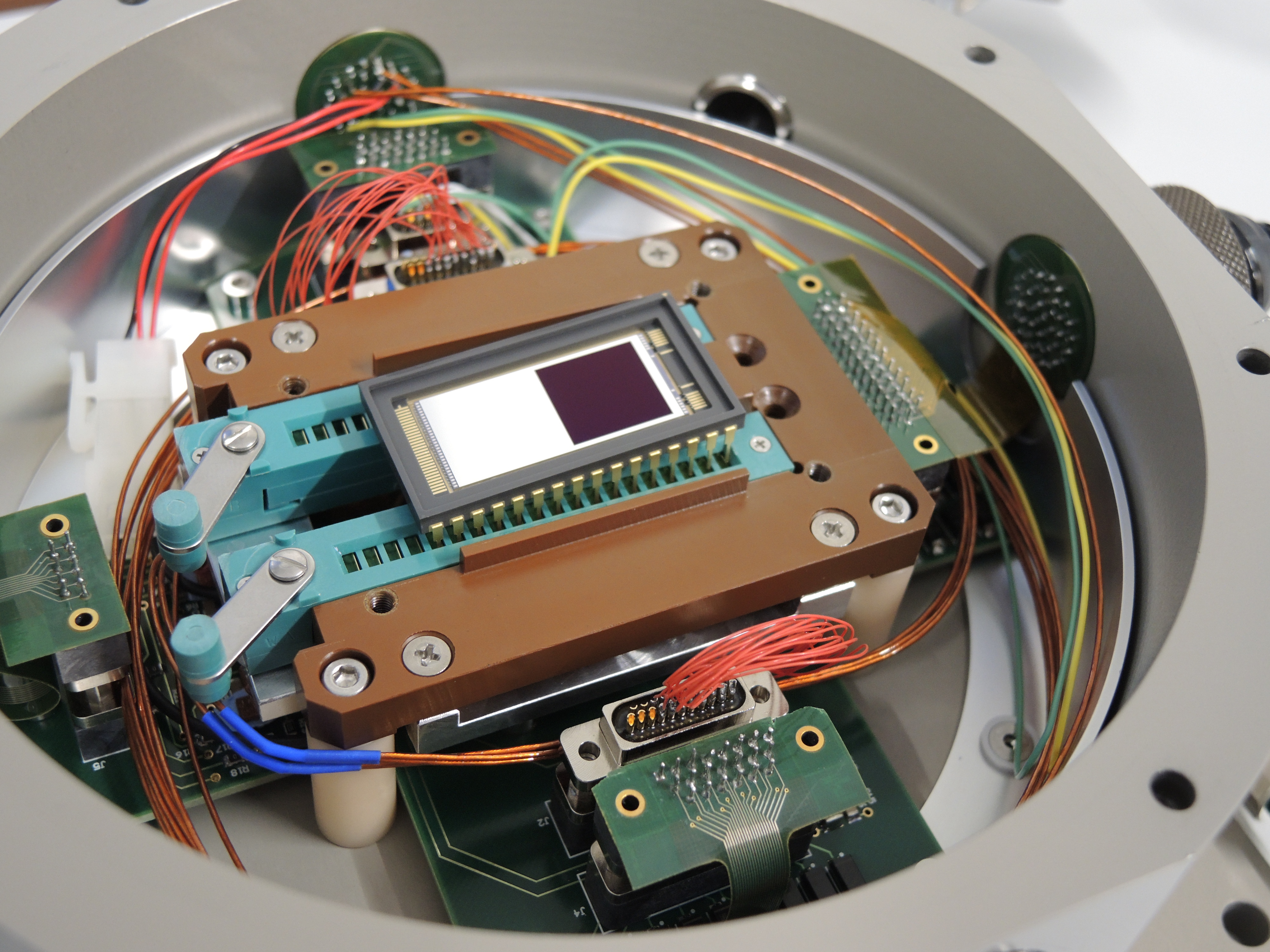}
\caption{CHEOPS CCD mounted in the cryostat of the University of Geneva.}
\label{fig:CCD_cryo}      
\end{figure}

The CCD gain sensitivity depends upon temperature and operating voltage. Both dependencies were studied and characterised. In the case of temperature, it is in the order of \SI{1}{ppm \per \milli \K} of noise. The operating voltages dependency is slightly more complex and depends on which voltage is being looked at. Generally, the dependency of these contributors are between 5 and \SI{40}{ppm \per \milli \V} (see \cite{Deline:2020}) for details). This fact led to the need for ensuring very stable temperatures and voltages for the CCD during measurements which turned out to be a major challenge to the instrument. Predictions, on the CCD gain sensitivity contributions to the overall noise budget (together with the T-dependence of the quantum efficiency) based on calibration measurements, show that the contribution is of the order of \numrange{3}{10} ppm. For details on the instrument calibration, the reader is referred to \cite{Deline:2020}.

In summary, the SEM/FPM represents the camera of the CHEOPS instrument that is controlled by the BEE as a higher-level computer.

\subsection{Back End Electronics (BEE)}
\label{sebsec:BEE}
The BEE is the main computer of the instrument. On one hand, it provides power to the entire payload and data interfaces to the platform on-board computer, on the other hand, it interfaces with the SEM. Similarly to the SEM, it contains a Data Processing Unit (DPU) and a Power Supply and Distribution Unit that provides conditioned power for itself and the SEM. The DPU hardware is based on the GR712, which contains two LEON3 processors and provides the space wire interface to the SEM and MIL-1553 interfaces towards the spacecraft. The DPU mass memory from 3D-Plus provides a FLASH memory in the configuration of 4 Gbit times eight-bit. For the effective operation of the processor, four components are used to provide 32-bit access and error detection and correction. The BEE is hosting the main instrument software, the In-Flight Software (IFSW), which provides a high-level control of the SEM and FPM. Additionally, its main functions are to perform the data handling, the centroiding of the stellar image that is used by the AOCS, and the data compression.

\section{Flight Software and Data Processing}
\label{sec:IFSW}
This section describes the basic observation sequence and IFSW functionality. The data processing is generally performed by the IFSW running on the BEE but some processing (e.g. windowing) is already performed by the SEM.

\subsection{Nominal observations}
A normal science observation by CHEOPS will always use the same automated on-board procedure called ``nominal science''. Once the instrument is in the appropriate mode (``pre-science"), for which the CCD is switched on and stabilised to operational temperature together with the front-end electronics, the main instrument computer parameters are updated for the specific observation, and the S/C has slewed to the target direction, the following steps will be performed:

\begin{enumerate}
    \item \textbf{Target acquisition}
    
The instrument acquires full-frame images, identifies all the stars in the region of interest (ROI), and performs pattern matching against the uploaded star map pattern using the angle method. A second, magnitude-based, algorithm is used for bright stars. Once located, the measured offset between the target star and the line-of-sight of the telescope is communicated to the AOCS system and a pointing correction is made. 
    
    This process continues until the location offset is smaller than the pre-defined target acquisition threshold or until a maximum number of iterations has been reached. Once the target is successfully acquired, the IFSW makes an autonomous transition to the next step. For more details on the target acquisition see \cite{Loeschl:2016,Loeschl:2018}.
    
    \item \textbf{Calibration frame 1}
    
    The telescope takes one full-frame image\footnote{Multiple images can be configured for special observations.}. This image is sent to the ground to characterise the stellar field.
    
    \item \textbf{Science observation}
    
    The instrument observation mode is changed to ``window mode'' and images are taken with a cadence equal to the exposure time (except for exposure times shorter 1.05 s). For each image, the instrument computes the centroid in a configurable ROI at the center of the sub-window (default \SI{51x51} {px}) and communicates the position offset to the AOCS system. Note that after step 1 the star should be located very close to the center of the sub-window. 
    
After the requested number of window images have been acquired, the instrument makes the transition to “pre-science” mode.\footnote{A post-science 2nd calibration frame can be configured.}
\end{enumerate}

\subsection{On-board Science Data Processing}
\label{subsec:data_proc}
CHEOPS is observing one star at a time intending to download all the acquired raw images without on-board data processing. However, the exposure time determines the cadence at which subsequent images are being taken and therefore ultimately sets the total amount of data to be downloaded. This total amount may occasionally exceed the daily available data down-link rate of \SI{1.2}{\giga b \per \day} making on-board data processing unavoidable. 

\begin{figure*}[htb]
  \includegraphics[width=0.95\textwidth]{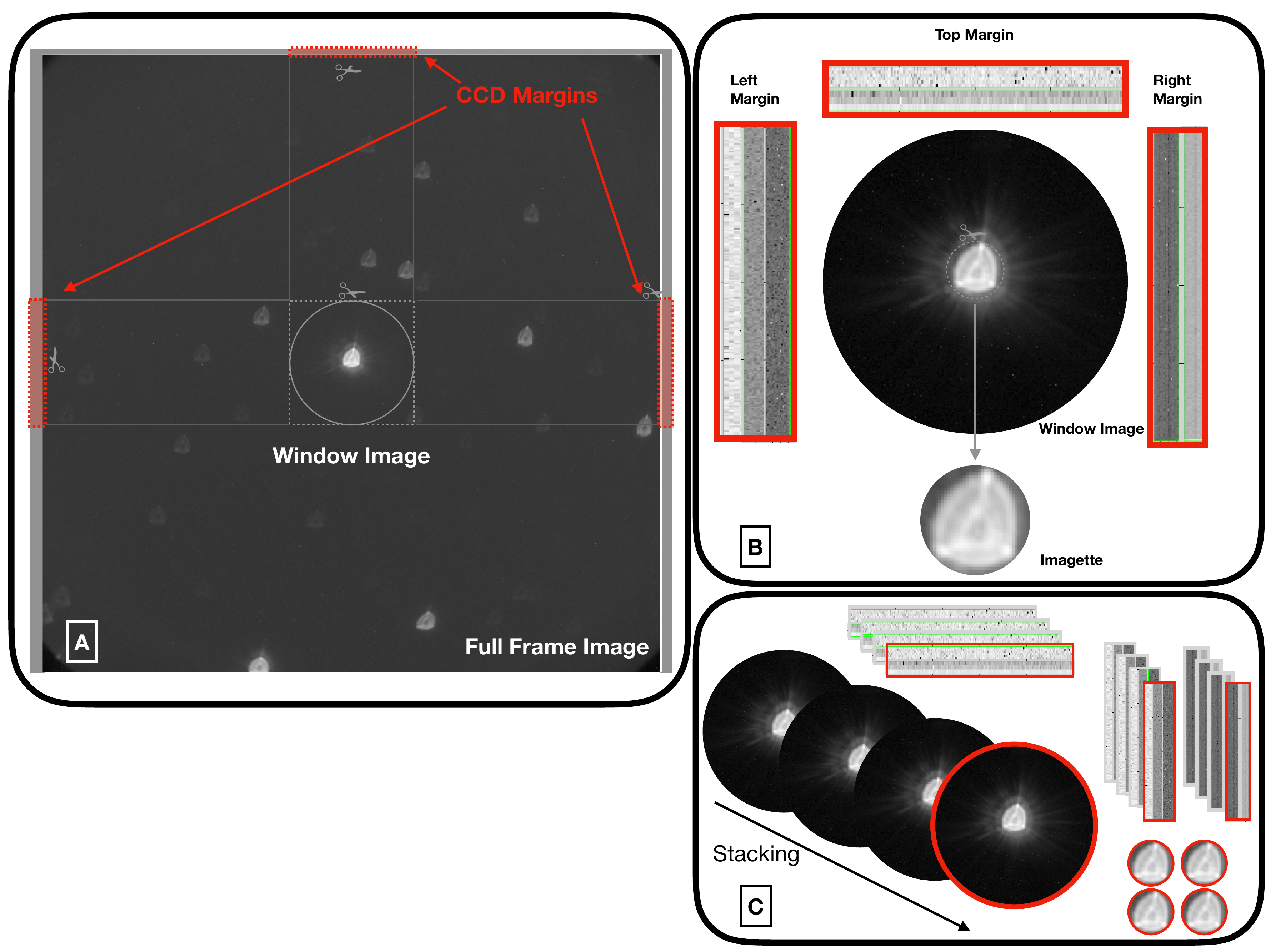}
\caption{This figure illustrates the on-board data processing. Panel A shows a full-frame image (1024x1024 plus CCD margins). Due to the limited bandwidth capabilities, all full-frame images cannot be down-linked to the ground. Therefore, circular window images with the target on their centre are cropped. The same is done to the corresponding section of the margins (details in panel B). If the exposure time of the image is longer than 30 seconds, all window images and margins are sent to the ground without any further on-board manipulation. However, if the exposure time is shorter than 30 seconds images have to be stacked on-board. In that case, small ``imagettes'' containing only the PSF of the target star, are also cropped but they are down-linked to the ground without stacking. For example, as illustrated in panel C, if the exposure time is 15 seconds, one stacked window image resulting from co-adding four images will be down-linked, together with the corresponding stacked margins and the four ``imagettes''. }
\label{fig:SDP}       
\end{figure*}

To reduce the total amount of data to be down-linked to the ground, we crop the full-frame images and generate circular ``window images'' of \SI{200} {pxs} diameter centred on the target star as shown in Figure ~\ref{fig:ccd}. For images taken with roughly 30 seconds or longer exposure times, no further data reducing action is necessary and all the raw ``window images" can be downloaded at a cadence equal to the exposure time. For images taken with exposure times shorter than 30 seconds, we reduce the total data amount by stacking on-board all individual images acquired within 60 seconds by co-adding them pixel-by-pixel. Only the stacked images are downloaded. For example, if the exposure time is \SI{15} {seconds}, four images will be stacked on-board and the resulting staked image down-linked at a cadence of one stacked image every 60 seconds.

To mitigate the loss of information associated with this process, small ``imagettes'' of radius  \SI{25} {pxs}, centred on the target star, are cropped from the individual ``window images" before stacking and are down-linked individually. This is done to facilitate, for example, the correction of cosmic ray hits inside the PSF. Figure~\ref{fig:SDP} sketches the on-board handling of the science data: panel A represents the full-frame image and the elements that are cropped out of it; panel B shows the circular window image, the margins' structure (see Sec. \ref{OTA}) and the ``imagette'' that is extracted in case stacking is necessary; and panel C shows the data structure that is down-linked in case of stacking. Note that if no stacking is performed ``imagettes'' are not extracted as they are not needed.
 
The sequence of steps performed on-board on the science images can be summarised as follows: 
 
\begin{enumerate}
    \item A circular window image centred on the target star is cropped from each image acquired.
    \item The CCD margins (top overscan and top dark, right and left darks, right and left blanks, and side overscan) are cropped from each image. They correspond to the column/row position of the window image. 
\end{enumerate}
If the exposure time is shorter than \SI{20}-\SI{30}{s} (final numbers are still being evaluated), the stacking of images is required and the following steps are performed.
\begin{enumerate}[{2}a.]
    \item The pixel within a small region centred on the target stars are extracted. These ``imagettes" are preserved at a full cadence (i.e. equal to the exposure time).
    \item The windowed images and corresponding margins are co-added pixel-by-pixel.
\end{enumerate}
Finally, the following steps are performed in all cases.
\begin{enumerate}\setcounter{enumi}{2}
    \item The (stacked) image, the (stacked) side margins, the individual  ``imagettes'' (if stacking was performed) plus additional housekeeping information (temperatures, voltages, etc.) are collected in a file and compressed using arithmetic compression. This file is called the ``compression entity''. \change{The compression achieved depends upon the entropy of the image, which is a function of numerous factors such as, for example, the number of stars in the background, exposure time, size of the image, etc. In practise, the compression factors actually measured, range between 2.5 and 3.3.}
    \item This compression entity is sent to the SC mass memory via PUS service 13. 
\end{enumerate}
The detailed treatment of all side margins, imagettes, etc. is highly configurable and can be adjusted to match available bandwidth and data reduction needs.

\section{Model philosophy and Verification Approach}
\label{sec:model_philosophy}
The CHEOPS verification approach was defined to remain compatible with the programmatic boundaries applicable to the project. The main drivers for the definition of the CHEOPS verification approach were the schedule constraint (spacecraft had to be ready for launch by end 2018) and had to allow for independent verification of instrument and platform  to the largest possible extent. Also, a complete qualification and acceptance cycle was required for the instrument, taking into account the new design and limited heritage at unit level, while the platform was benefiting from a considerable level of flight heritage. 

The platform equipment underwent an early confirmation of qualification against the CHEOPS environment or definition of any required punctual delta qualification, while the platform functional test specification and test procedures are based on AS250 product line, with adaptations for CHEOPS specifics. The instrument functional verification at spacecraft level was focused on the verification of the interfaces with the platform and functional testing. Instrument optical performances were not verified at satellite level, but regular health checks performed during the satellite test campaign.
Accounting for the drivers defined above, the following model philosophy was established:
\begin{itemize}
   \item CHEOPS Structural Qualification Model: composed by the Platform Structural Qualification Model  and the Instrument Assembly Structural and Thermal Model.
   \item CHEOPS Electrical and Functional Model:  composed by the Platform Electrical and Functional Model and the Instrument Assembly Engineering Model (EM or EQM -electronic units).
   \item CHEOPS Electrical Hybrid Model: composed by the Platform Proto-Flight Model  and the Instrument Assembly Electrical Qualification or Engineering Model.
   \item CHEOPS Proto-Flight Model: composed by the Platform Proto-Flight Model and the Instrument Assembly Proto-Flight Model.
\end{itemize}

This approach enabled an independent verification of the instrument and the platform as stand-alone elements, complemented by regular interface tests. The instrument models set the timing for the satellite verification, while the platform models were ready with enough margins to maintain the tight schedule. The availability of the instrument EMs and EQMs (electronic units) proved essential to anticipate integration activities and to test interfaces ahead of the availability of flight units. To support the verification of the Radio Frequency compatibility between the spacecraft and the CHEOPS ground segment, a representative radio frequency suitcase, including the Honeywell S-band Transceiver EM, was used to perform Radio-Frequency Compatibility Tests  with all foreseen ground stations.

\section{The Ground Segment }
\label{sec:Ground_segment}
The CHEOPS Ground Segment is composed of the Mission Operation Centre (MOC) located at Torrej\'on de Ardoz (INTA, ES), the Science Operation Centre (SOC) at University of Geneva (CH) and two Ground Stations  located respectively at Torrej\'on (main station) and Villafranca (backup station) (ES). The MOC and SOC are national contributions from the Mission Consortium (respectively from Spain and from Switzerland).  

All operations, including Launch and Early Orbit Phase (LEOP) and In-Orbit Commissioning (IOC), have been executed from the MOC. ESA has provided the Mission Control System and the spacecraft simulator as part of the satellite procurement contract and also provides Collision Avoidance services through the European Space Operation Center (ESOC) located in Darmstadt (Germany). The Mission Planning System, driven by essentially time-critical science observation requirements, has been provided by the Consortium and is part of the SOC. As for the other elements of the mission, the operational concept of CHEOPS reflects the fast-track and low-cost nature of the mission, following two basic principles: 1) maximum reuse of existing infrastructure and operational tools, 2) high levels of both on-board autonomy and automation in the operations, to minimise the required manpower. 

The responsibility for LEOP and in-orbit commissioning lay with ESA as the mission architect. During LEOP, the Kiruna and Troll ground stations were made available by ESA to complement the Torrej\'on and Villafranca stations, providing additional passes and enabling an early acquisition of the spacecraft after the separation from the launcher. The nominal duration of the LEOP was 5 days, concluding when the spacecraft was safely in the nominal operational orbit and ready for starting payload operations. A 2.5 months IOC phase has followed the launch with nominal science operations having started on April 18, 2020.

After commissioning, the responsibility for the mission operations were handed over to the Consortium with an ESA representative following the operational aspects. ESA will remain directly involved in critical decisions, such as for example collision avoidance manoeuvres and the de-orbiting of the CHEOPS satellite at the end of the mission.

\subsection{Mission Operations Centre (MOC)}
\label{subsec:MOC}
The \change{MOC} is hosted and operated by the Instituto Nacional de T\'ecnica Aeroespacial (INTA), located in Torrej\'on de Ardoz near Madrid, Spain. 

\begin{figure}
    \centering
    \includegraphics[width=0.95\textwidth]{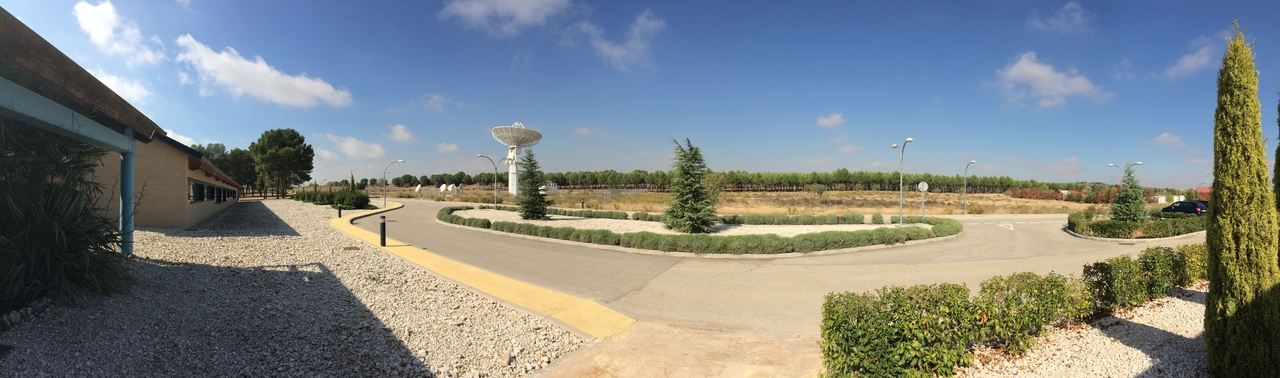}
    \caption{The CHEOPS Mission Operations Centre located at INTA near Madrid, Spain.}
    \label{fig:GS_moc_outside}
\end{figure}

The MOC responsibilities for CHEOPS mission include Mission operations and Spacecraft disposal at end of mission, including:
\begin{itemize}
   \item Mission control planning rules; 
   \item Implementation of the activity plan; 
   \item Spacecraft and instrument control and monitoring; 
   \item Orbit and attitude determination and control; 
   \item Management of failure and anomalies;
   \item Spacecraft deactivation (decommissioning phase);
   \item Operation of ground station for all mission phases, plus a respective backup capability.
\end{itemize}

The MOC test and validation philosophy reflects the small mission character of the mission by keeping the level of documentation minimal and demonstrations of the capabilities by simulations, tests, etc. For the same reasons, it has been automated to a large extent to limit the operation costs. 

\subsection{Science Operations Centre (SOC)}
\label{subsec:SOC}
The Science Operations Centre (SOC) is under the responsibility of the University of Geneva and is physically located at the Department of Astronomy of the University. Seven additional institutes or industrial partners from the CHEOPS Mission Consortium countries have made important contributions to the SOC. 

The ultimate objective of the CHEOPS Science Operations Centre (SOC) is to enable the best possible science with the CHEOPS satellite by delivering the following elements:
\begin{itemize}
    \item Mission planning after spacecraft commissioning;
    \item Science operations visibility, reporting, and knowledge management;
    \item Support to the CHEOPS science team and, in exceptional cases, indirect support to guest observers;
    \item Science operations system study, design, and requirements;
    \item Mission planning tool development;
    \item Quick look analysis to monitor the performance of the instrument;
    \item Development and maintenance of the data reduction pipeline to process raw data and deliver science ready data products (images and light-curves);
    \item Mission data archiving;
    \item Science data distribution.
\end{itemize}

The CHEOPS operational concept is based on a weekly cycle where the SOC generates the sequence of activities to be executed on-board and the MOC uplinks at once the corresponding commands to the spacecraft for a week of autonomous in-flight operations. Typically, around 30 individual targets are observed weekly for durations ranging from one hour to one week (median visit duration is about 8 hours).

From a mission planning perspective, the key element resides in the stringent time-critical nature of most observations that have to be scheduled (transits and occultations). To maximise the scientific return of the mission, the SOC uses a genetic algorithm to optimise the short- and long-term planning of observations. The goodness of these schedules is evaluated with a merit function that accounts primarily for the scientific priority of the planned observations and the GTO/GO time balance, and to a lesser degree for the filling factor, the on-source time, and the completion rate of individual observation requests. An overall filling factor in excess of 95\% is readily obtained with this approach.

The Data Reduction Pipeline \cite{Hoyer:2020} is run automatically once triggered by the processing framework. There is no interaction with external agents and there is no interactive configuration of the pipeline. The complete processing can be separated in 3 main steps: 1) the calibration module which corrects the instrumental response, 2) the correction module in charge of correcting environmental effects and 3) the photometry module which transforms the resulting calibrated and corrected images into a calibrated flux time series or light curve (see \cite{Hoyer:2020}). Each of these modules consists of successive processing steps which are run sequentially as the output of one step is used as an input of the next one. In addition to the reduced light curves and their associated contamination curve, the calibrated and corrected images, the pipeline generates a visit processing report. This report allows the user to get direct insight into the performance of each step of the data reduction.

A Project Science Office (PSO) has been established to serve as an interface between the Science Team and the SOC to, for example, verify and adapt to the proper format, the target list, and other information passed to the SOC. The PSO also provides the instrument reference files derived from the observations of the Monitoring and Characterisation Programme. The PSO is also the interface to ESA for the Announcement of Opportunities under the responsibility of ESA. The PSO provides support to the AOs.

Finally, an Instrument Team tracks the contributions to the noise budget and establishes the Instrument Operations Plan. Derived from the data taken during the on-ground calibration activities the Instrument Team will provide a set of instrument reference files that will be mainly used for the Data Reduction inside the SOC. An updated set of instrument reference files will be provided at the end of the IOC when the initial instrument performance was assessed in-flight. The operational role of the Instrument Team is the resolution of instrument anomalies and to implement changes of the on-board software resulting from the follow-up activities or following requests from the SOC or the PSO.

\section{Launch and Early Orbit Phase (LEOP)}
\label{leop}
At 8:54:20 UT on December 18, 2019, CHEOPS was successfully launched as a secondary passenger on a Soyuz-Fregat rocket from Kourou in French Guiana (see Fig. \ref{fig:cheops-liftoff}). The primary payload in terms of mass was the first of the second generation of Cosmo-SkyMed dual-use radar reconnaissance satellites for the Italian government. The 2.2-metric-ton satellite, manufactured by Thales Alenia Space for the Italian Space Agency, separated from the Fregat upper stage 23 minutes after launch. CHEOPS with its wet mass of 273-kilogram entered its intended near-polar dusk-dawn Sun-synchronous orbit following separation 2 hours 24 minutes after launch. Finally, three additional CubeSats were also on-board as auxiliary payloads. EyeSat, a 3U CubeSat (5 kg) student satellite, ANGELS, a 30-kilogram technology miniaturisation test satellite (both launched for CNES), and ESA’s OPS-SAT which will be testing and validating new techniques in mission control and on-board satellite systems.
\begin{figure}[htb]
    \centering
    \includegraphics[width=0.65\textwidth]{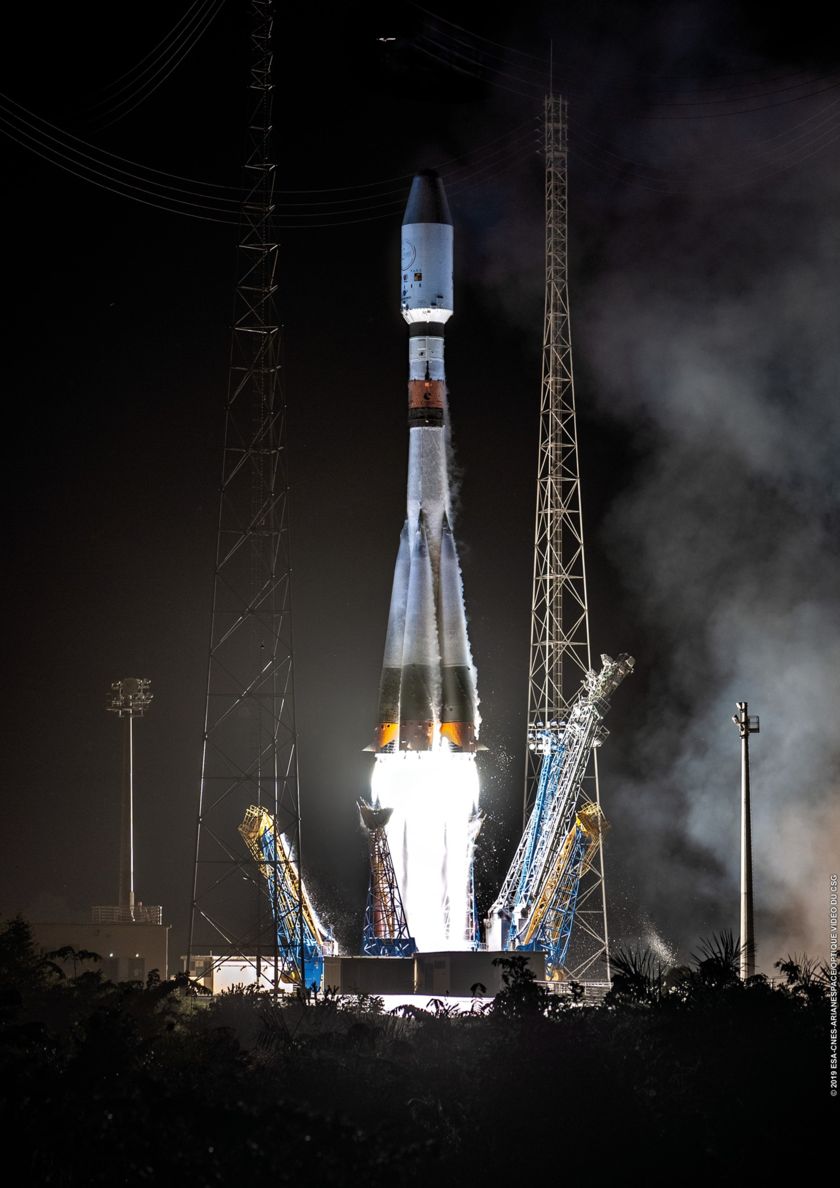}
    \caption{Successful CHEOPS launch on a Soyouz-Fregat from Kourou (French Guyana) at 8:54 UT on December 18, 2019}
    \label{fig:cheops-liftoff}
\end{figure}

The execution of LEOP and IOC operations, under ESA responsibility as Mission Architect, were delegated to the platform contractor Airbus Defence \& Space Spain. All activities have been performed from the mission operation centre at INTA near Madrid (Spain) supported by the Troll (Antarctica) and Kiruna (Sweden) ground stations in addition to the stations of Torrej\'on and Villafranca in Spain.

At 11:43 UT on day 1, the first telemetry acquisition by the Troll Ground Station in Antarctica arrived at the mission operation centre as planned. This started 4.5 days of activities aiming at ensuring that the satellite can be put into safe mode waiting for in-orbit commissioning activities to start early January. As a first measure to prevent condensation from early out-gassing, the temperature of the focal plane assembly was increased above the nominal operation value. Two-way Doppler measurements for orbit determination were started showing that the error in orbit injection was less than \SI{300}{m} in the semi-major axis. Star trackers were started and configured and overall power convergence achieved. On day 2, the normal mode of the attitude and orbit control system was achieved and the main survival-redundant equipment checks performed while the pyrovalves for orbit control manoeuvres opened. On day 3, all ground stations and satellite equipment were declared healthy and with prime units in use. The execution of the calibration orbit control manoeuvre was successfully done. The automation system for INTA ground stations (Torrej\'on and Villafranca) and for the CHEOPS operation centre was tested. On day 4, the satellite was in its final operational orbit after 3 orbit control manoeuvres (one for calibration and two for correction). The difference in semi-major axis between the targeted and achieved orbit is less than 30m. No additional manoeuvres for orbit maintenance will be needed during CHEOPS's lifetime. Finally, on day 5 all the nominal units as well as all the redundant units critical for the survival of the satellite were declared working and healthy. A total of \SI{198}{g} of propellant was used during this phase. 

LEOP was completed on 22 December without recording major anomalies and with all subsystems using the nominal equipment chain. The spacecraft was declared ready for in-orbit commissioning and put in safe mode on December 22, 2019, with only basic spacecraft maintenance taking place until the start of the activities at the mission operation centre on January 7, 2020.

\section{In-Orbit Commissioning (IOC)}
\label{ioc}
The IOC activities started at the MOC on January 7, 2020. Teams from ESA, Airbus, INTA, the University of Bern and Geneva were assembled at MOC during the first few weeks of these activities which were divided in four phases:
\begin{enumerate}
    \item IOC-A: Instrument switch-on (January 7 - January 29, 2020)
    
    This first phase was dedicated to switching on both the nominal and redundant chains of the instrument. Further activities were dedicated to confirm the thermal control performances of the instrument. Furthermore, the calibration of the dark current and the CCD pinning curves were performed while the cover of the instrument was still closed. This phase was completed on January 29 with the successful opening of the cover. 
    
    \item Monitoring \& Characterisation and AOCS performances verification (January 30 - February 26, 2020)
    
    After opening the cover, the first images were taken without the payload in the tracking loop and the offset between the telescope’s line-of-sight and the star trackers measured and corrected. Detailed calibration of star tracker’s optical heads with respect to the spacecraft’s reference frame was performed based on instrument data but without enabling payload measurements to enter the AOCS control loop in order to allow for an accurate pointing even without instrument measurements. Subsequently, observations with the Payload in the Loop for tracking the target star have been performed. This has led to a full characterisation of the pointing abilities of the system. Several additional detailed characterisation were performed during this phase aiming at defining the point-spread function (see Sect.~\ref{subsec:psf}), the extent of stray light as a function of angle of incidence, the amount of dark current, the number of hot pixels, the measure of the gain stability, as well as an update of the location of the South-Atlantic Anomaly.
    
    \item IOC-C: Nominal observations (February 27 - March 8, 2020)
    
    This phase was dedicated to actual observations aimed at verifying the science performance requirements of the system (see Sec.\ref{sec:requirements}). Some illustrative results are shown in Sec. \ref{subsec:photometric_precision}. Monitoring and characterisation activities were continued during this phase. 

    \item IOC-D: End-to-end operational validation (March 9 - March 25, 2020)
    
    This phase was dedicated to the final validation of the entire chain of operations by carrying out activities as they would be in routine operations. This included the first observations of a few exoplanet transits. Finally, activities for the preparation of the hand-over of the operational responsibility to the Consortium have been conducted.  
   
\end{enumerate}

While overall the IOC activities went rather smoothly, not surprisingly there were many issues that arose which required analysis and additional measurements. As an example, one can mention the surprising measure of significant stray light on the detector even though the cover of the telescope was still closed. In this configuration, only dark images were expected as the optical cavity was supposed to be completely shielded from the outside. After considerable investigation and many additional measurements with different telescope pointings, the root cause could be identified. The small hole in the cover that was used to illuminate the CCD one last time before shipping the satellite to Kourou to verify that the detector was still working was leaking even though it had been taped closed. It was concluded that either the tape fell off during launch or that it was more transparent than expected in the infrared. As no leaks from anywhere else in the system could be detected, the problem was no longer relevant after the opening of the cover. 

A more difficult issue, which is still present, was the evidence of a much larger number of hot pixels than expected based on the COROT data (of order 70\% more). Also, most of these hot pixels being telegraphic, their correction is somewhat more difficult. To reduce their number, the operating temperature of the CCD has been reduced in several tests from \SI{233}{\K} to \SI{228}{\K} and finally to \SI{223}{\K}. Because the latter lower temperature might create temperature stability issues in some extreme pointings and because all calibrations had been carried out at \SI{233}{\K}, it was decided to operate the CCD at the intermediate temperature of \SI{228}{\K} which resulted in a decrease of hot pixels by a factor 3. 

\subsection{Point-Spread Function (PSF)}
\label{subsec:psf}
As mentioned in Sect.~\ref{OTA}, the telescope has been deliberately defocused to mitigate jitter in the satellite pointing and the saturation of pixels for bright stars. A key activity during IOC-B was the exact measurement of the actual size and shape of the PSF defined as the region where 90\% of the total energy received from the star in form of light is being deposited. The detailed knowledge of the PSF is of importance to ensure the required photometric performances. Figure~\ref{fig:psf} shows the actual CHEOPS PSF as measured during IOC-B. 

\begin{figure}[htb]
\centering
  \includegraphics[width=0.85\textwidth]{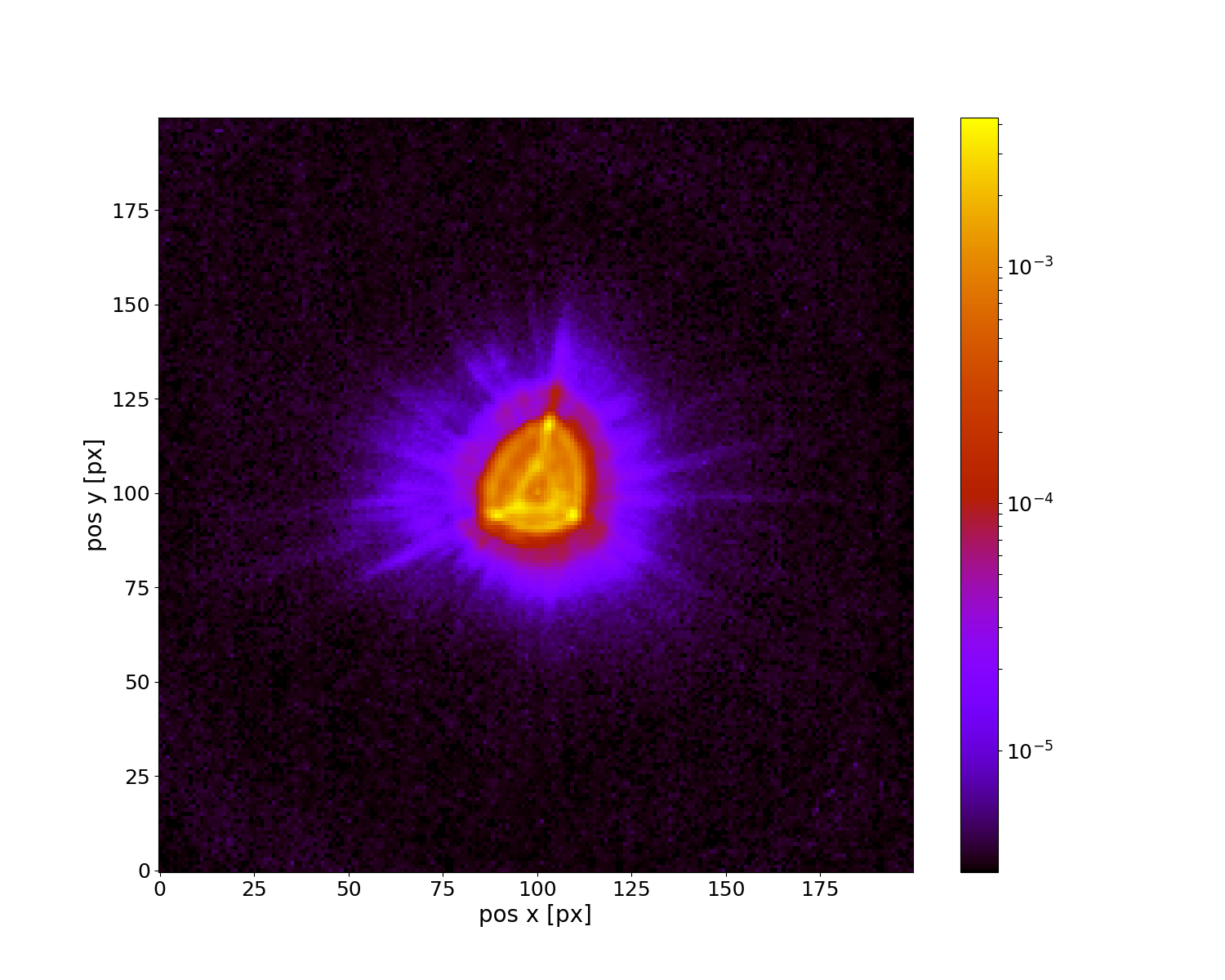}
\caption{CHEOPS has a de-focused PSF where \SI{90}{\%} of the total energy is inside a radius of 16 pixels. In this figure the PSF flux distribution in white light is shown as measured during the in-orbit commissioning.}
\label{fig:psf}     
\end{figure}

The measured PSF shows an expected triagonal deformation which originates from the strain on the mirror stemming from its three-point fixation mechanism. Such a mechanism was used to ensure that the telescope could withstand high loads as the launch vehicle was not known at the time of the design. To some extent some engineering choices have been driven by the stability of the final PSF rather than by its symmetry. Being now measured in the absence of gravity, the PSF appears, as expected, significantly more symmetric than the one measured during calibration in the laboratory. It was also noted that the size of the PSF of \SI{16}{pxs} is larger than the expected range of \SIrange{12}{15}{pxs}. Finally, following the predictions obtained by comparing thermoelastic models \cite{Magrin:2018} with the laboratory measurements, the actual PSF is much smoother than the ones recorded on the ground. This smoothness coupled with the spreading of the light over a larger area translate in a factor 4.5 reduction in the intensity of the brightest spots of the PSF as compared to the laboratory measurements. A larger and smoother PSF has two consequences depending upon the magnitude of the target stars. For bright stars it is extremely positive as it reduces the risk of pixel saturation while for faint stars it makes distinguishing signal from noise more difficult. \change{As a consequence, for very faint targets,  hot pixels or cosmic rays can lead to degradation in the pointing accuracy when using the instrument as a fine guidance sensor. As the pointing of the payload without this feedback loop is very accurate on its own, it was decided no to use the payload in the loop for stars fainter than magnitude 11.} 

\subsection{Photometric Precision and Stability}
\label{subsec:photometric_precision}
As mentioned in Sect.~\ref{subsec:photometric_accuracy}, the requirement for bright stars (science requirement 1.1) called for a photometric precision of \SI{20} {ppm} (goal: \SI{10} {ppm}) in 6 hours of integration time for G5 dwarf stars with V-band magnitudes in the range \(6\leq V\leq 9\,\)mag. To illustrate the verification of this requirement, we show the light curve obtained over 47 hours of observation of HD 88111, a magnitude $V=9.2$ and effective temperature $T_{eff}=5330$ K star for which GAIA \cite{GAIADR2:2018} provides a radius of $0.9 R_{\odot}$. This star was chosen as a well known stable star and hence ideally suited to verify the achievable precision. The exposure time was 30 seconds without stacking images and the photometry was obtained using a circular aperture of \SI{30}{px} in radius. 

\begin{figure}[htb]
  \includegraphics[width=1.0\textwidth]{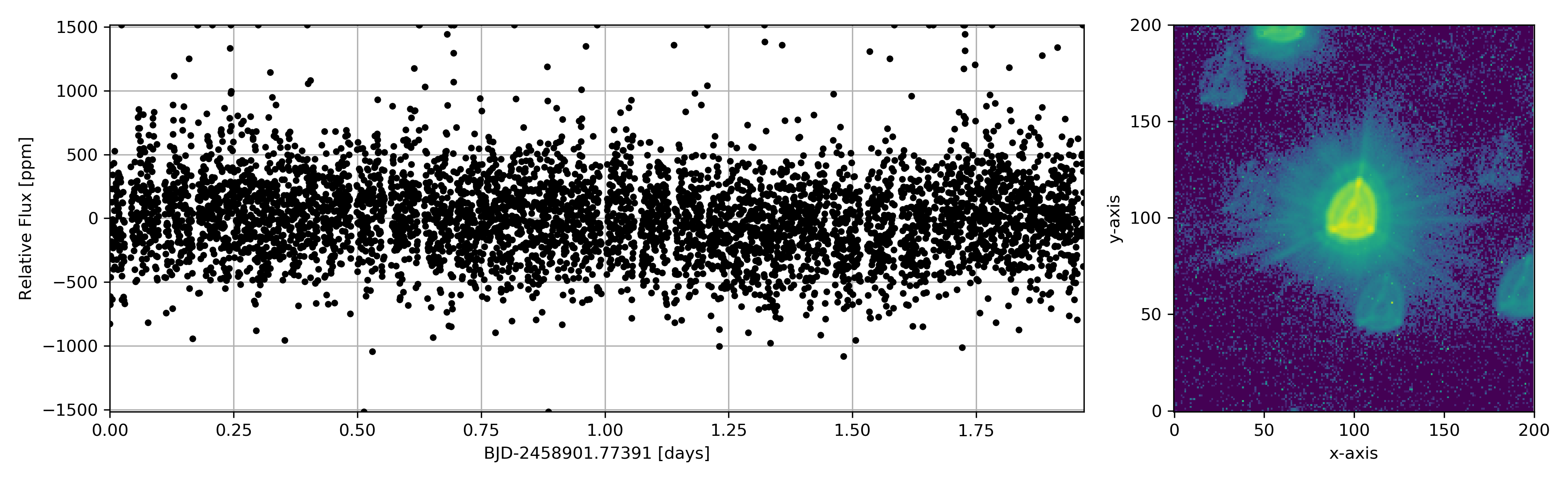}
\caption{Left: Photometric light-curve of the star HD 88111 obtained from images taken every 30 seconds over a period of 47 hours without any detrending. The gaps in the data are due to the Earth occulting the target star and to the passage of the spacecraft through the South Atlantic Anomaly, a period during which the data are discarded. Over a period of 6 hours, the precision achieved is \SI{15.5}{ppm}; Right: One of the image of HD 88111 as recorded by CHEOPS}
\label{fig:photometric_stability_bright}     
\end{figure}

\change{The photometric precision and stability is estimated by finding the transit depth that can be detected with a signal-to-noise ratio of 1. This is essentially the same method used to calculate the Kepler combined differential photometric noise value \cite{Christiansen:2012}}. For a six hour period of observation, the achieved photometric precision is \SI{15.5}{ppm}; well within the precision requirement outlined in Sect.~\ref{subsec:photometric_accuracy}. \change{A similar precision is obtained after analysis of any six hour period during the 47 hours of observation. This precision is achieved without any detrending and therefore reflects the intrinsic stability of CHEOPS.}

The star TYC 5502-1037-1 was chosen to test the faint end of the photometric precision of CHEOPS. This is a $V=11.9$ magnitude star, with an effective temperature of $T_{\rm eff}=4750$ K and a radius of $R=0.7 R_{\odot}$. Estimating the photometric precision of this observation was not as straight forward as for HD 88111. The first observation made was inadequate for a precision analysis as the window location was chosen too close to one of the margins of the CCD and a hot pixel appeared in the PSF during the visit. A second 3 hour observation was made later on. \change{ A 75 ppm precision was achieved for this observation, which makes this case compliant with science requirement 1.2. Note that, as in the case of HD 88111, no detrending of the data was performed.}

\change{In summary, measurements taken during commissioning demonstrate that CHEOPS meets the photometric precision requirements on both the bright and faint stars. The determination of the actual detailed photometric performances of CHEOPS is ongoing work 
that will be carried out based on the analysis of actual science targets by the CHEOPS Science and Instrument Teams over the coming months. }

\change{Finally, even though stray light was no issue in the results reported in this paper, its effects were carefully studied by means of dedicated observations. The function measuring the rejection of stray light, the point source transmission (PST), could be estimated by observing close to the Moon and was found to be in the range $10^{-9}$ to $10^{-12}$ for incoming light with an angle of incidences greater than 35 degrees. These measured values are within the error bars of simulations of the optical system carried out early on during the design phase of the instrument. It is worth mentioning that unexpected stray light was detected in some images taken along a line of sight close to the un-illuminated Earth limb. After analysis, this was attributed to atmospheric glow. Since this effect is unpredictable it is nearly impossible to correct, and affected images have to be discarded.}

\subsection{KELT-11b: Radius determination of a bloated planet}
\label{subsec:light-curve}
During IOC-D a few stars known to host planets were targeted by CHEOPS as part of the end-to-end validation of the operational process. The giant planet KELT-11b was among these targets. Discovered by the KELT survey in 2016 \cite{Pepper:2017}, the radius of the planet measured by transit photometry is $R_p=1.37_{-0.12}^{+0.15} ~ R_{Jup}$ while its mass derived from radial velocity data is $M_p=0.195 \pm 0.18 ~ M_{Jup}$. The planet orbits the evolved subgiant star HD 93396 of magnitude $V=8$ in a period of $ P = 4.736529 \pm  0.00006$ days. As evidenced by these data, KELT-11b is a so-called bloated giant planet having about 20\% of the mass of Jupiter but a radius larger by almost 40\%. 

To analyse the data we used {\tt pycheops}\footnote{\url{https://github.com/pmaxted/pycheops}} version 0.7.0. {\tt pycheops} has been developed specifically for the analysis of CHEOPS data and uses the qpower2 algorithm \cite{Maxsted:2019} for fast computation of transit light curves. Optimisation of the model parameters is done using {\tt lmfit}\footnote{\url{https://lmfit.github.io/lmfit-py/}} and {\tt emcee} \cite{Foreman-Mackey:2013} is used to sample the posterior probability distribution  (PPD) of the model parameters.
 
The observed data comprise 1500 flux measurements in a photometric aperture with a radius of \SI{29}{arcsec} from images with exposure times of \SI{30}{s} covering the transit of KELT-11\,b on 2020-03-09. We excluded 8 exposures that provided discrepant flux measurements and 101 exposures observed in a narrow range of spacecraft roll angle for which there is excess scatter in the flux ($\sim 200$\,ppm) caused by scattered moonlight. Our model for the observed flux, $f(t)$ is of the form $$f(t) = F(t;\vec{c}) ~ T(t; D, W, b, T_0, h_1, h_2)~ S(t;S_0,\omega_0) + \epsilon(\sigma_w),$$ where F(t;\vec {c}) is a scaling factor, $T(t; D, W, b, T_0, h_1, h_2)$ is the transit model computed using qpower2, and $S(t; S_0,\omega_0)$ is a model of the intrinsic stellar variability. The random noise $\epsilon(\sigma_w)$ is assumed to be white noise with a variance $\sigma_i^2 + \sigma_w^2$, where $\sigma_i$ is the error bar on a flux measurement $f(t_i)$ provided by the CHEOPS data reduction pipeline. The vector of ``de-trending'' coefficients $\vec{c}$ is used to compute a linear model for instrumental effects, $ F(t;\vec{c}) =  \vec{B}\cdot\vec{c}$, where the matrix of basis vectors $\vec{B}$ includes the estimated contamination of the photometric aperture using simulated images of the CHEOPS field of view ({\tt lc\_contam}), plus 6 functions of the form $\sin(n \phi)$ and $\cos(n \phi)$ where $n=1,2,3$ and $\phi$ is the spacecraft roll angle. The stellar noise is assumed to have a power spectrum of the form $S(\omega) = \sqrt{2/\pi}\,S_0/[(\omega/\omega_0)^4+1]$. The likelihood for a given set of model parameters $\{\theta\}$, ${\cal L}(f(t); \{\theta\}; S_0,\omega_0,\sigma_w)$, is calculated using {\tt celerite} \cite{Foreman-Mackey:2017}. The parameters of the model for the transit at time $T_0$ of a star with radius $R_{\star}$ by a planet of radius $R_{\rm pl}$ in a circular orbit of semi-major axis $a$ and inclination $i$ are $D = (R_{\rm pl}/R_{\star})^2 = k^2$ (depth), $W=(R_{\star}/a)\sqrt{(1+k)^2 - b^2}/\pi$ (width), and  $b = a\cos(i)/R_{\star}$ (impact parameter). We fixed the value of the orbital period for this analysis. We assume uniform priors on $\cos(i)$, $\log(k)$ and $\log(a/R_{\star})$. The stellar limb darkening is modelled using the power-2 law with parameters $h_1 = 0.715 \pm 0.011$ and $h_2 = 0.442 \pm 0.05$ taken from \cite{Maxsted:2018}. 

Next to the transit signal, we observe stellar variability with an amplitude of approx. 200 ppm, correlated over timescales between 30 minutes and 4 hours. We attribute these variations to the effects of stellar granulation, as their amplitude and frequency behaviour is in excellent agreement with the empirical relations derived from Kepler data \cite{sulis:2020}. We model them using the GP described above, with the following priors imposed on the hyper-parameters: $log S_0 = -23.4 \pm 0.6, log \, \omega_0 = 5.6 \pm 0.3$ and $log \, \sigma_w = -10.4 \pm 3.7$. These values were found from a fit to the residuals from an initial least-squares fit. The time scale and amplitude of the stellar variability for models with these priors is similar to that seen in other stars a similar spectral type to KELT-11 \cite{Kallinger:2014}.
    
The best-fit model for these KELT-11 data is shown in Fig.~\ref{kelt11fig} and the model parameters with their standard errors (calculated from the mean and standard deviation of the PPD) are given in Table~\ref{kelt11table}. The RMS residual from the fit is 198\,ppm.

\begin{figure}
\centering
\includegraphics[width=0.95\textwidth]{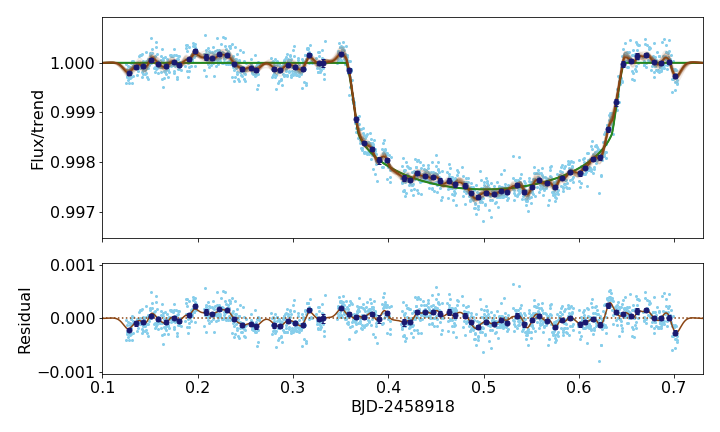}
\caption{Observed light curve of KELT-11 and model fit from {\tt pycheops}. The measured fluxes (light blue points) are also shown binned in time (dark blue points). The transit model (green line, barely visible) is shown in the upper panel together with the several realisations of our complete model including stellar noise sampled from the PPD. The lower panel shows residuals from the transit model plus instrumental effects (blue points) together with the best-fit stellar noise model (brown line).}
\label{kelt11fig}%
\end{figure}

\begin{table}
\caption[]{Parameters of our {\tt pycheops} model fit to the transit of KELT-11\,b observed by CHEOPS. Parameters of the form $df/dx$ are de-trending coefficients. The basis vectors are normalised so that value of these coefficients corresponds to the amplitude of the instrumental noise correlated with the corresponding parameter. The mean stellar density is $\langle\rho_{\star}\rangle$.}
\label{kelt11table}
\begin{center}
  \begin{tabular}{@{}lrr}
\hline
 \noalign{\smallskip}
\multicolumn{1}{@{}l}{Parameter} &
 \multicolumn{1}{l}{Value} &
 \multicolumn{1}{l}{Error}  \\
\noalign{\smallskip}
\hline
\noalign{\smallskip}
$T_0$ ~~[BJD$_{\rm TDB}$] & 2458918.5017 & $\pm 0.0007 $ \\
$D$    & 0.0021  & $\pm 0.0001$ \\
$W$    & 0.0614  & $\pm 0.0003$ \\
$b$    &     0.4 & $\pm 0.2$ \\
$h_1$  &    0.71 & $\pm 0.01$ \\
$h_2$  &    0.45  & $\pm  0.05$ \\
$df/d({\tt lc\_contam})$~~ [ppm] &   203 & $\pm  100$ \\ 
$df/d\sin(\phi)$ ~~[ppm]  &$    20 $& $\pm 27$ \\
$df/d\cos(\phi)$ ~~[ppm]  &$   -24 $& $\pm 30$ \\
$df/d\sin(2\phi)$ ~~[ppm] &$    35 $& $\pm 22$ \\
$df/d\cos(2\phi)$ ~~[ppm] &$    23 $& $\pm 23$ \\
$df/d\sin(3\phi)$ ~~[ppm] &$   -40 $& $\pm 22$ \\
$df/d\cos(3\phi)$ ~~[ppm] &$   -49 $& $\pm 18$ \\
$\log S_0$      & $-23.3 $ &$ \pm 0.3$ \\
$\log \omega_0$ & $ 5.6 $ & $\pm 0.1$ \\
$\log \sigma_w $ &  $ -10.4 $& $ \pm  1.5$ \\
\noalign{\smallskip}
\multicolumn{3}{@{}l}{Derived parameters} \\
 k            &   0.0463  & $\pm 0.0003$ \\
$a/R_{\star}$ &   5.0     & $\pm 0.1$ \\
$\log (\langle\rho_{\star}\rangle)$~~ [solar units] & $ -1.12$ & $\pm 0.03$\\
\noalign{\smallskip}
\hline
\end{tabular}
\end{center}
\end{table}
With the value for $k=0.0463$ and its uncertainty $0.0003$ as shown in Table~\ref{kelt11table} and adopting a value for the radius of the star KELT-11 of $R_{\star}=2.807 \pm 0.036 \, R_\odot$ we obtain a radius for the planet KELT-11b (using $R_\odot=696342\,km$) of $R_{\rm pl} = k \, R_{\star} = 90500 \pm 1747 \, km$ which translates (taking $R_{Jup}=69911 \, km$) into $R_{\rm pl} = 1.295 \pm 0.025 \, R_{Jup}$. This value is consistent with the value obtained earlier \cite{Pepper:2017} \change{but with an error bar 5 times smaller}. We note that the error bar provided includes contributions from both the error associated with the measurement of $k$ and from the one associated with  $R_{\star}$: $\Delta R_{\rm pl} = \left(\partial R_{\rm pl} / \partial k\right)\Delta k + \left(\partial R_{\rm pl} / \partial R_{\star} \right)\Delta R_{\star} = 0.00839 R_{Jup} + 0.0167 R_{Jup} = 0.025 R_{Jup}$. This shows that it would be still possible to significantly reduce the error on the planet radius by improving on the estimates of the stellar properties.

\section{Summary and Conclusions}
\label{summary_conclusions}
CHEOPS is the first small-class mission in ESA's science programme. As such, it is a demonstrator showcasing the ability of ESA and its Member States to develop fast, low-cost science missions. As the first of its kind, CHEOPS had to pave the way in many respects in order to maintain the aggressive schedule and remain within budget. At the end of 2018, after only 6 years from the initial mission proposal selection, 4 years from the satellite PDR  and 2.5 years after the satellite CDR, the fully integrated CHEOPS satellite has completed all planned tests. The satellite QAR was passed in February 2019, marking flight readiness. Finally, CHEOPS was successfully launched as a secondary passenger on a Soyuz-Fregat rocket from Kourou (French Guiana) in December 2019. End of March 2020, in-orbit commissioning was over and the satellite declared to meet all its requirements. The responsibility for the routine science operations was handed over to the CHEOPS Consortium by ESA. The mission reached this milestone within schedule and budget. 

As of April 18, 2020, the routine science operations including Guaranteed Time Observations and Guest Observations have begun. Early results support what was already surmised from the commissioning results, namely that CHEOPS is indeed the precision photometric machine it was designed to be. 

\begin{acknowledgements}
CHEOPS is an ESA mission in partnership with Switzerland with important contributions to the payload and the ground segment from Austria, Belgium, France, Germany, Hungary, Italy, Portugal, Spain, Sweden, and the United Kingdom. The CHEOPS Consortium would like to gratefully acknowledge the support received by all the agencies, offices, universities, and industries involved. Their flexibility and willingness to explore new approaches were essential to the success of this mission. 

The Swiss participation to CHEOPS has been supported by the Swiss Space Office (SSO) in the framework of the Prodex Programme and the Activit\'e Nationales Compl\'ementaires (ANC), the Universities of Bern and Geneva as well as well as of the NCCR PlanetS and the Swiss National Science Foundation. The early support for CHEOPS by Daniel Neuenschwander is gratefully acknowledged. 

In Austria, the CHEOPS activities have been supported by the ESA Prodex programme, the Austrian Research Promotion Agency (FFG), and the Austrian Academy of Sciences (\"{O}AW).

The Belgian participation to CHEOPS has been supported by the Belgian Federal Science Policy Office (BELSPO) in the framework of the PRODEX Program, and by the University of Liege through an ARC grant for Concerted Research Actions financed by the Wallonia-Brussels Federation. MG is F.R.S.-FNRS Senior Research Associate. V.V.G. is a F.R.S.-FNRS Research Associate

The team at LAM acknowledges the CNES support, including grant 124358, and the support of the Direction Technique of INSU with P.G. assignment.

L.L.K. and G.M.S were supported by the Hungarian National Research, Development and Innovation Office (NKFIH) grant GINOP-2.3.2-15-2016-00003
and the Hungarian Academy of Sciences. G.M.S also acknowledges the NKFIH K-119517 grant.

For Italy, CHEOPS activities have been supported by the Italian Space Agency, under the programs: ASI-INAF n. 2013-016-R.0 
and ASI-INAF n. 2019-29-HH.0.

In Portugal, this work was supported by FCT - Funda\c{c}\~ao para a Ci\^encia e a Tecnologia through national funds and by FEDER through COMPETE2020 - Programa Operacional Competitividade e Internacionaliza\c{c}\~ao by these grants: UID/FIS/04434/2019; UIDB/04434/2020; UIDP/04434/2020; PTDC/FIS-AST/32113/2017 \& POCI-01-0145-FEDER-032113; PTDC/FIS-AST/28953/2017 \& POCI-01-0145-FEDER-028953; PTDC/FIS-AST/28987/2017 \& POCI-01-0145-FEDER-028987. S.C.C.B. and S.G.S. acknowledge support from FCT through FCT contracts nr. IF/01312/2014/CP1215/CT0004, IF/00028/2014/CP1215/CT0002. O.D.S.D. is supported in the form of work contract (DL 57/2016/CP1364/CT0004) funded by national funds through Funda\c{c}\~ao para a Ci\^encia e Tecnologia (FCT).

We also acknowledge support from the Spanish Ministry of Science and Innovation and the European Regional Development Fund through grants ESP2016-80435-C2-1-R,  ESP2016-80435-C2-2-R,  PGC2018-098153-B-C31, PGC2018-098153-B-C33, ESP2017-87676-C5-1-R,  and MDM-2017-0737 Unidad de Excelencia Mar\'{i}a de Maeztu- Centro de Astrobiolog\'{i}a (INTA-CSIC),  as well as the support of the Generalitat de Catalunya/CERCA programme. The MOC activities have been supported by the ESA contract No. 4000124370.

Finally, XB, SC, DG, MF and JL  acknowledge their role as ESA-appointed CHEOPS science team members.

\end{acknowledgements}

%
 \section*{Conflict of interest}

 The authors declare that they have no conflict of interest.

\bibliographystyle{spmpsci}      
\bibliography{CHEOPS.bib}   

\end{document}